\documentclass[%
superscriptaddress,
 amsmath,amssymb,
 aps,
prd,
twocolumn,
]{revtex4-1}

\usepackage{graphicx}
\usepackage{xcolor}
\usepackage{ulem}
\usepackage{dcolumn}
\usepackage{bm}
\usepackage[mathlines]{lineno}
\usepackage{hyperref}
\usepackage{upgreek,textcomp}
\usepackage{float}

\begin{document}

\preprint{APS/123-QED}
 
\title{Probing gluon distributions with $D^0$ production at the EicC}

\author{Daniele Paolo Anderle}
\email{dpa@m.scnu.edu.cn}
\affiliation{Key Laboratory of Atomic and Subatomic Structure and Quantum Control (MOE),
Institute of Quantum Matter, South China Normal University, Guangzhou 510006, China}
\affiliation{Guangdong Provincial Key Laboratory of Nuclear Science, Institute of Quantum Matter,
South China Normal University, Guangzhou 510006, China}
\affiliation{Guangdong-Hong Kong Joint Laboratory of Quantum Matter,
Southern Nuclear Science Computing Center, South China Normal University, Guangzhou 510006, China}

\author{Aiqiang Guo}
\email{guoaq@impcas.ac.cn}
\affiliation{Institute of Modern Physics, Chinese Academy of Sciences, Lanzhou, Gansu Province 730000, China}
\affiliation{University of Chinese Academy of Sciences, Beijing 100049, China}

\author{Felix Hekhorn}
\email{felix.hekhorn@mi.infn.it}
\affiliation{Tif Lab, Dipartimento di Fisica, Università di Milano and\\ INFN, Sezione di Milano, Via Celoria 16, I-20133 Milano, Italy}

\author{Yutie Liang}
\affiliation{Institute of Modern Physics, Chinese Academy of Sciences, Lanzhou, Gansu Province 730000, China}
\affiliation{University of Chinese Academy of Sciences, Beijing 100049, China}
\affiliation{Guangdong Provincial Key Laboratory of Nuclear Science,\\ Institute of Quantum Matter, South China Normal University, Guangzhou 510006, China}

\author{Yuming Ma}
\affiliation{Institute of Modern Physics, Chinese Academy of Sciences, Lanzhou, Gansu Province 730000, China}

\author{Lei Xia}
\affiliation{University of Science and Technology of China, Hefei, Anhui Province 230026, China}

\author{Hongxi Xing}
\affiliation{Key Laboratory of Atomic and Subatomic Structure and Quantum Control (MOE),
Institute of Quantum Matter, South China Normal University, Guangzhou 510006, China}
\affiliation{Guangdong Provincial Key Laboratory of Nuclear Science, Institute of Quantum Matter,
South China Normal University, Guangzhou 510006, China}
\affiliation{Guangdong-Hong Kong Joint Laboratory of Quantum Matter,
Southern Nuclear Science Computing Center, South China Normal University, Guangzhou 510006, China}
\affiliation{Southern Center for Nuclear-Science Theory (SCNT), Institute of Modern Physics, Chinese Academy of Sciences, Huizhou, Guangdong Province 516000, China}

\author{Yuxiang Zhao}
\affiliation{Institute of Modern Physics, Chinese Academy of Sciences, Lanzhou, Gansu Province 730000, China}
\affiliation{Southern Center for Nuclear-Science Theory (SCNT), Institute of Modern Physics, Chinese Academy of Sciences, Huizhou, Guangdong Province 516000, China}
\affiliation{University of Chinese Academy of Sciences, Beijing 100049, China}
\affiliation{Key Laboratory of Quark and Lepton Physics (MOE) and Institute of Particle Physics, Central China Normal University, Wuhan 430079, China}

\date{\today}


\begin{abstract}
The Electron-Ion Collider in China (EicC) has been proposed to study the inner structure of matter and fundamental 
laws of strong interactions. In this paper, we will present a conceptual design of the tracking system
based on the state-of-art silicon detector and Micro-Pattern Gaseous Detector at the EicC and 
demonstrate that it will enable us to reconstruct charm hadron with good significance, hence study gluonic parton distribution functions in 
nucleons and nuclei, as well as gluon helicity distributions. The impact study using reweighting techniques shows
that the impact of the EicC will be mainly in the large $x$ region. It complements similar physics programs at the Electron-Ion Collider at Brookhaven National Laboratory.

\end{abstract}
                         
\maketitle

\tableofcontents

\section{Introduction}\label{sec:intro}
Understanding the visible world we are living in is of fundamental importance to modern sciences. The building blocks of visible matter, nuclei, and nucleons, are in turn composed of quarks and gluons bound together by a strong force. The
underlying theory describing the strong interactions between quarks and gluons is 
Quantum Chromodynamics (QCD). To gain a deeper insight into the inner structure of 
nucleons, an Electron-ion collider in China (EicC) has been proposed~\cite{Anderle:2021wcy}. It will provide highly 
polarized electrons and protons with a variable center-of-mass (c.m.) energies ranging from 15 to 20 GeV and
with a luminosity of $(2-3) \times 10^{33} \,\textrm{cm}^{-2} \cdot \textrm{s}^{-1} $. A series of unpolarized ion 
beams ranging from Carbon to Uranium will also be available at the EicC.

The EicC will explore a wide range of physics topics: the spin structure of nucleons, the partonic
structure of nuclei, the parton interaction with the nuclear environment, and exotic states, especially
those with heavy flavor quark contents. In order to achieve such broad physics goals, a hermetical detector 
system will be constructed with cutting-edge technologies. In this paper, we discuss the requirements 
for charged-particle tracking and describe a conceptual design of a tracking system based on 
state-of-the-art silicon pixel detectors coupled with Micro-Pattern Gaseous Detectors (MPGD).
Based on the high precision of the vertexing and momentum resolution for the tracking system, a feasibility study with quantitative estimations on the measurements of $D^0$/$\bar{D}^0$
production at the EicC is presented. 
Measurements of charm hadron production
in deep-inelastic scattering (DIS) can provide constraints on the gluon distributions as the relevant unpolarized/polarized charm structure functions provide direct
access to the respective gluon distributions from the leading order, through the photon-gluon
fusion (PGF) process.

These constraints can complement the extraction of gluon distributions via QCD evolution or higher-order corrections in fully inclusive DIS.
Similar studies have already been performed for the Electron-Ion Collider (EIC) at Brookhaven National Laboratory~\cite{Kelsey:2021gpk,Anderle:2021hpa,Dong:2022xbd,AbdulKhalek:2021gbh}.
Our study here shows that the EicC can make a significant impact on the unpolarized or polarized gluon distributions of
nucleons and nuclei through charm hadron production
in the large-$x$ region ($x > 0.1$), which is complementary to the studies at the EIC. 

The rest of the paper is organized as follows: Section \ref{sec:tracking} describes the proposed tracking system as well as its 
performance and Section \ref{sec:physics} presents the physics goals and simulation studies on the determination of gluon distributions with $D^0$/$\bar{D}^0$ production in the DIS process. It contains both unpolarized gluon distribution studies
with unpolarized e+p and e+Au collisions and polarized gluon distribution study with longitudinally
polarized e+p collisions. In Section \ref{sec:summary} we summarize our conclusions.

\section{A tracking detector conceptual design at the EicC}
\label{sec:tracking}

\subsection{Requirements}
In the EicC baseline design, an electron beam energy is 3.5 GeV, and proton beam energy of 20 GeV, resulting in a c.m. energy of 16.7 GeV and a cross-section of 20.8 $\mu b$. The luminosity is expected to be $L \rm = 2\times 10^{33}\, cm^{-2}s^{-1}$ with an interaction rate of 83.2 kHz. The PYTHIA \cite{Sjostrand:2006za} simulation shows the final state particles are concentrated near pseudo-rapidity $\eta$ = 1, with a particle density rate of $dN/d\eta dt = \rm 8\times10^4/s$.

Among the final state particles, the scattered electrons provide crucial information to most of the physics programs, in particular those focusing on the processes of DIS, semi-inclusive deep inelastic scatterings (SIDIS), deeply virtual Compton scattering (DVCS), and so on. It is noticed that with increasing $Q^2$, the scattered electrons are less boosted to negative pseudo-rapidity. For physics requiring $Q^2$ larger than 1 $\rm{GeV^2}$ (such as SIDIS or DVCS), a detector coverage of $\eta>-2$ is sufficient for the scattered electron. In addition to scattered electrons, the other final state particles are also important and have been studied. The number of final state pions is about 1–2 orders of magnitude larger than those of kaons and anti-protons. The momenta of hadrons in other $\eta$ regions are also investigated. At a pseudo-rapidity smaller than 1, the momenta of final state hadrons are expected to be smaller than 6 GeV, while at large pseudo-rapidity region ($\eta>2$), the final state hadrons have momenta up to 15 GeV. Thus, the detection and particle identification (PID) of the final state hadrons needs to be considered at various pseudo-rapidity regions.

As a high luminosity machine, EicC could reduce the statistical uncertainty down to a few percent for many measurements. To cope with the small statistical uncertainty, we need a matching systematic uncertainty, which requires a good detector. For example, a tracking detector with a tracking resolution of a few percent is necessary. A momentum resolution of 1\% for the central coverage and 2\% for small angles are marked according to the experiences from similar experiments. For the first version of the conceptual design, we divide the EicC detector into the central detector and the forward (backward) detector. The central detector consists of the barrel part and two endcaps, and it will be constructed inside a solenoid. In this paper, we focus on the central detector.

\subsection{The magnetic field}{\label{sec_B_field}}

The magnetic field, which is provided by a superconducting solenoid, plays an essential role in momentum measurement for the charged tracks. 
\begin{figure}[H]
\centering
\includegraphics[width=0.48\textwidth]{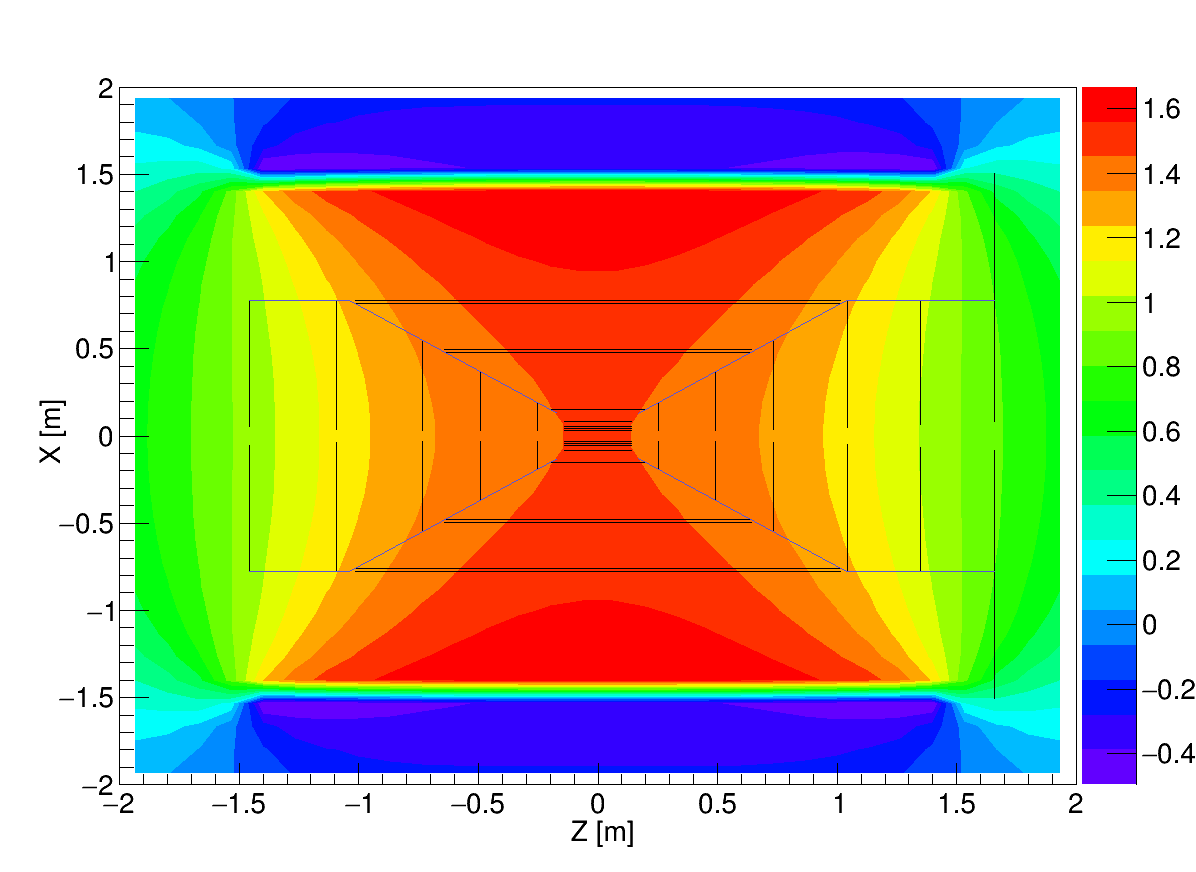}
\caption{The calculated Bz of EicC solenoid (in Tesla). The black lines show the detector schematic. }
\label{fig:Bfield}
\end{figure}
The superconducting solenoid has a maximum field strength of 1.5 T with an inner bore radius of 1.5 m and a coil length of 3.0 m. A stronger magnetic field certainly is beneficial for the momentum resolution, meanwhile, it also decreases the tracking efficiency for the low momentum track due to the larger bending capability. From the PYTHIA simulation, the transverse momentum of the majority of the track is less than 2 GeV, therefore, a field strength of 1.5 T is at the balance point between the momentum resolution and tracking efficiency. The 2D map of the field is also calculated as shown in Fig.~\ref{fig:Bfield}. And it is sufficient to provide the barrel tracking with a uniform high-field region, while also containing the PID devices and the Electromagnetic Calorimeter (ECal).

\subsection{Choice of technology}

To satisfy the requirements of EicC physics programs, we investigated various tracking technologies and detector designs, including all silicon tracking systems based on the Monolithic Active Pixel Sensors (MAPS) and a hybrid pixel detector and MPGD tracking system. The latter one is proven to have better performance and is utilized in this paper. We will explain why these technologies are chosen for the first conceptual design.  

MAPS provides high granularity, low power consumption, and consequently low material budget, as well as the required readout speed in one device. Therefore, it is considered the best detector technology to satisfy the requirements of the EicC vertex and tracking detector. In addition, The integration of charge collection and readout capabilities into one silicon substrate is well-suited for the required level of integration and acceptance coverage of the EicC. The MAPS detector has been developed for three generations. The first generation was deployed in the STAR Heavy Flavour Tracker\cite{Contin:2017mck} and in the ALICE ITS2~\cite{ALICE:2013nwm}. The latter used the ALPIDE sensor\cite{AglieriRinella:2017lym}, which is fabricated in a commercial 180 nm Complementary Metal Oxide Semicondutor (CMOS) imaging process provided by Tower Jazz. It provides better charge collection properties, radiation hardness, and signal processing capabilities compared to the traditional MAPS. The second generation of MAPS was designed to reach the rate and radiation tolerance capability typically required by high luminosity particle physics experiments. Recently, a large community is gathering to develop a third-generation MAPS in a 65 nm CMOS imaging technology for future experiments through the ALICE ITS3 project~\cite{Musa:2703140} and the CERN EP R\&D program. The ITS3 project is aiming to develop a three-layer vertex detector with an extremely low material budget. It will apply low power design techniques, large area, 2D stitched sensors thinned below 50 $\rm \upmu m$, thus it can be bent around the beam pipe to minimize cooling, support structure, and services in the active area, therefore, enabling a material budget of only 0.05\% $X_0$. Such a detector concept is a very attractive solution for the EicC vertex layers where an extremely low material budget coupled with the sensor’s high granularity will deliver the required vertex resolution. For the tracking layers and disks, a reticule-size version of the ITS3 sensor will be developed with a more conventional design of support structures.

MPGD technologies such as Gas Electron Multiplier (GEM)~\cite{Sauli:1997qp}, Micro Mesh Gaseous Structures (Micromegas)~\cite{Giomataris:1995fq}, Resistive Micro Well (µRWELL)~\cite{Bencivenni:2014exa} are widely used for tracking in various particle physics experiment across the world. These technologies typically combine a gaseous device for electron amplification with high granularity strips or pads anode readout PCB to provide a combined excellent 2D space point resolution ($\sim 50\, \rm \upmu m$), fast signal per layer ($\sim$5\,ns), radiation hardness and large area capabilities at a significantly lower cost compared to silicon trackers.

EicC will adopt silicon MAPS near the interaction point and MPGDs farther out. This configuration allows for low-material budget tracking with sufficient redundancy over a large lever arm, which is critical to achieving the required momentum resolution. 

\subsection{Geometry}
\begin{figure}[htb]
\centering
\includegraphics[width=0.5\textwidth]{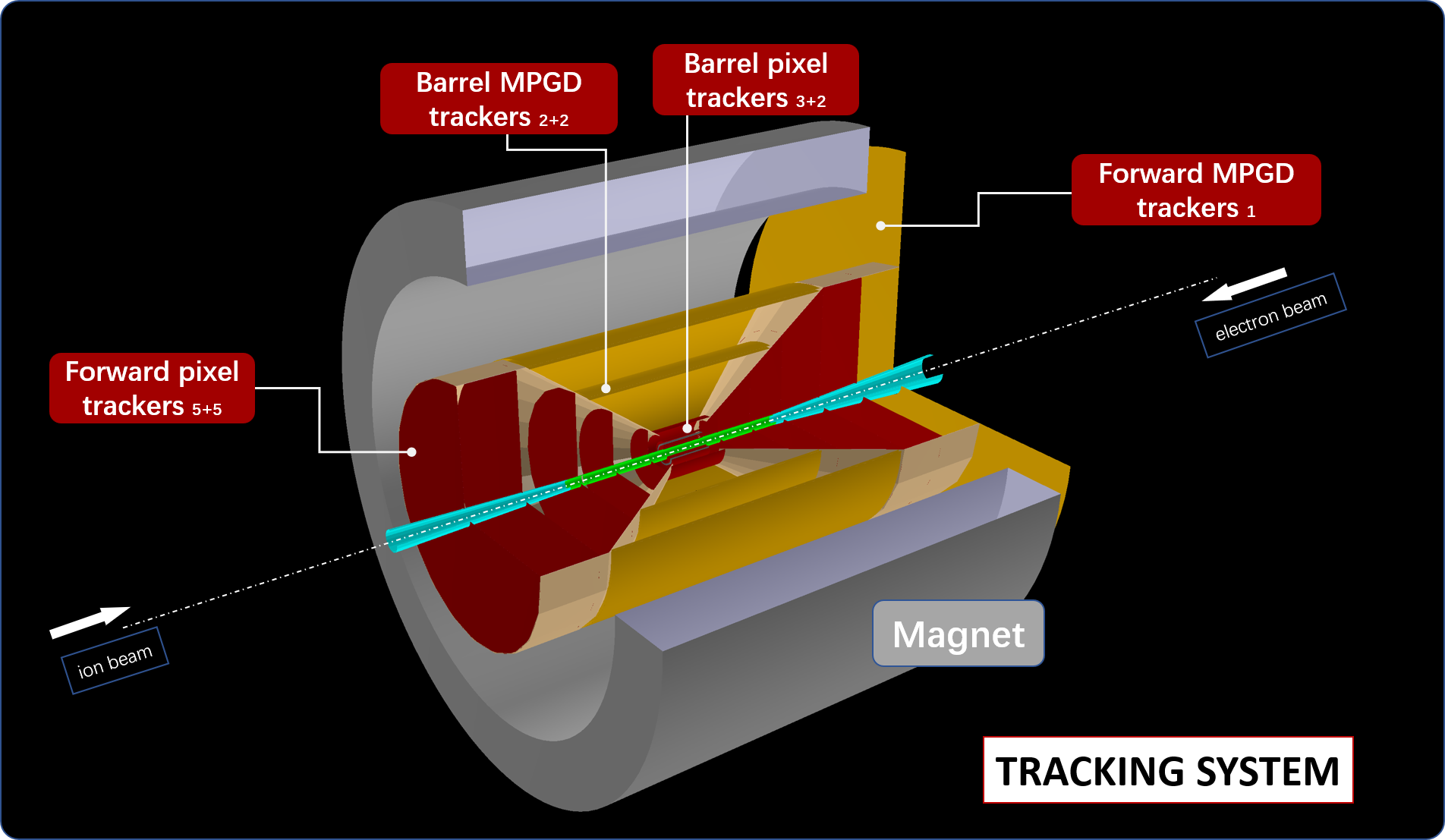}
\caption{The conceptual design of EicC tracking system. The silicon detctor, MPGD, and support structures are shown in the red, dark-yellow, and light-yellow components, respectively. The beryllium section of the beam pipe is shown in green, and the rest of the beam pipe is shown in cyan. The solenoid is shown in gray.}
\label{fig:EicC_tracker}
\end{figure}

The layout of the vertex and tracking system is illustrated in Fig.~\ref{fig:EicC_tracker}. The tracker coverage for low values of pseudorapidity ($|\eta|<1.1$) is utilized by a barrel hybrid tracking system including the inner silicon and outer MPDG layers. The inner silicon barrel consists of three vertex layers and two tracking layers. They occupy a region that has a maximum radius of 15\,cm and a total length of 38\,cm. The vertex layers deploy the wafer-scale, stitched sensors that are bent around the beam pipe, which is made of a beryllium cylinder with a radius of 3.17 cm. The tracking layers comprise the same stitched sensors but have different support structures. Two closely-spaced 2-D layers of Micromegas are chosen to cover the outermost barrel region. Their mean radii are approximately 48\,cm and 77\,cm, and their maximum total length of approximately 200 cm. The radii of each layer and their corresponding lengths along z are summarized in Table~\ref{tab:table_barrel}.

\begin{table}[htbp]
\caption{The radii (R), lengths, pitch(pixel) size, material budget ($X/X_0 (\%)$), and technology (Tech.) for the barrel tracking system}
\label{tab:table_barrel}
\begin{center}
\begin{tabular}{l|c|c|c|c}
\hline\hline
  R(cm)   &  Length(cm)  & Pitch Size ($\rm \upmu m$)  & $X/X_0 (\%)$ & Tech.    \\
\hline
    3.3  & 28.0 & 10 & 0.08 & Silicon \\
    4.4  & 28.0 & 10 & 0.08 & Silicon \\
    5.4  & 28.0 & 10 & 0.08 & Silicon \\
    8.0  & 28.0 & 10 & 0.08 & Silicon \\
    15.0  & 38.7 & 10 & 0.08 & Silicon \\
    47.7  & 127.5 & 150($\rm r\upphi$)$\times$150(z) & 0.40 & MPGD \\
    49.6  & 127.5 & 150($\rm r\upphi$)$\times$150(z) & 0.40 & MPGD \\
    75.6  & 202.0 & 150($\rm r\upphi$)$\times$150(z) & 0.40 & MPGD \\
    77.5  & 202.0 & 150($\rm r\upphi$)$\times$150(z) & 0.40 & MPGD \\
\hline
\end{tabular}
\end{center}
\end{table}

To cover the larger values of pseudorapidity ($|\eta|>1.1$), the tracking system consists of silicon disks followed by large-area Micromegas in the forward (proton/nucleus-going) directions and all silicon disks in backward (electron-going) direction. The silicon disks will use the same sensor technology as the vertex and tracking layers in the barrel region. There are five silicon disks in the both backward direction and the forward direction. They start from 25\,cm along the z direction on either side of the interaction point and extend to 145\,cm in the backward direction and 134\,cm in the forward direction. The minimum radii are determined by the divergence of the beam pipe while the maximum outer radius of the disks is approximately 77\,cm.  One Micromegas detector is implemented at 165\,cm in the forward direction from the IP. Its inner and outer radii are about 8\,cm and 150\,cm, respectively. In addition, it is located behind the dual ring imaging Cerenkov (dRICH) detector in the forward direction. This detector helps not only the momentum resolution in the forward direction but also the seed finding of the dRICH ring.  All the geometry parameters and position as well as the material budget are listed in Table~\ref{tab:table_endcap_pgoing} and Table~\ref{tab:table_endcap_egoing}.

\begin{table}[htbp]
\caption{The radii ($R_{in}$ and $R_{out}$), position (Z), pitch(pixel) size, material budget ($X/X_0 (\%)$), and technology (Tech.) for the end-cap region tracking system (proton going direction)}
\label{tab:table_endcap_pgoing}
\begin{center}
\begin{tabular}{c|c|c|c|c|c}
\hline\hline
 $R_{in}$ (cm)  & $R_{out}$ (cm) &  Z (cm)  & Pitch Size ($\rm \upmu m$)  & $X/X_0 \%$ & Tech.    \\
\hline
    3.2  & 18.6  & 25.0 & 10 & 0.08 & Silicon \\
    3.2  & 36.5  & 49.0 & 10 & 0.08 & Silicon \\
    3.5  & 54.7  & 73.0 & 10 & 0.08 & Silicon \\
    5.1  & 77.5  & 103.7 & 10 & 0.08 & Silicon \\
    6.6  & 77.5  & 134.3 & 10 & 0.08 & Silicon \\
    8.2  & 150.0  & 165.0 & 50($\rm r\upphi$)x250(r) & 0.40 & MPGD \\
\hline
\end{tabular}
\end{center}
\end{table}

\begin{table}[htbp]
\caption{The radii ($R_{in}$ and $R_{out}$), position (Z), pitch(pixel) size, material budget ($X/X_0 (\%)$), and technology (Tech.) for the end-cap region tracking system (electron going direction)}
\label{tab:table_endcap_egoing}
\begin{center}
\begin{tabular}{c|c|c|c|c|c}
\hline\hline
 $R_{in}$ (cm)  & $R_{out}$ (cm) &  Z (cm)  & Pitch Size ($\rm \upmu m$)  & $X/X_0 \%$ & Tech.    \\
\hline
    3.2  & 18.6  & -25.0 & 10 & 0.08 & Silicon \\
    3.2  & 36.5  & -49.0 & 10 & 0.08 & Silicon \\
    3.2  & 54.7  & -73.0 & 10 & 0.08 & Silicon \\
    4.0  & 77.5  & 109.0 & 10 & 0.08 & Silicon \\
    5.6  & 77.5  & 145.0 & 10 & 0.08 & Silicon \\
\hline
\end{tabular}
\end{center}
\end{table}

\begin{figure}[H]
\centering
\includegraphics[width=0.48\textwidth]{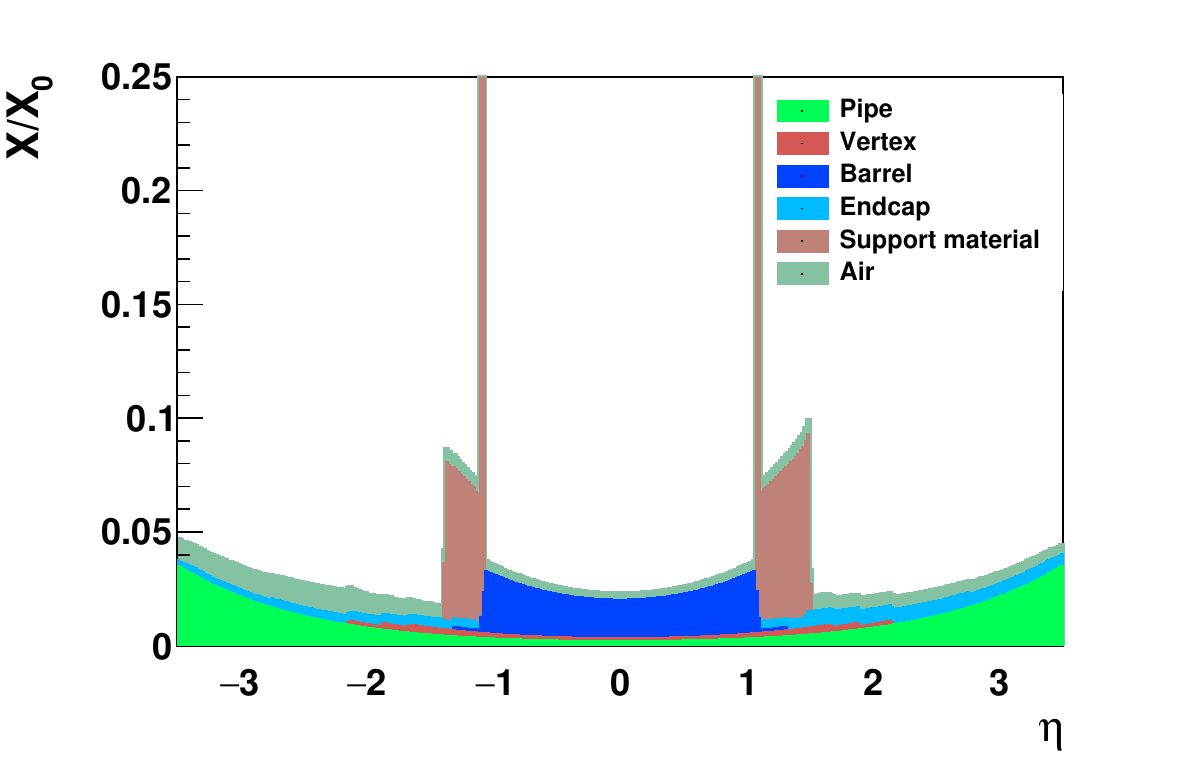}
\caption{The material budget as a function of pseudorapidity of EicC tracking system. The different colors show the contribution from the beam pipe (light green), vertex detector (dark brown), tracking doctor in the barrel part (blue) and end-cap part (cyan), the support structure (light brown), and the rest part (dark green).}
\label{fig:EicC_material}
\end{figure}

Both the barrel layers and the disks are made up of large-area, wafer-scale, stitched sensors based on ALICE-ITS3 design. 
Besides the active silicon volume, each layer includes components such as support structures between the layers, which combined correspond to an average material budget of 0.08\%\,$X_0$ per layer. Each Micromegas layer in the barrel and each of the forward disks has a material thickness well below 0.5\%\,$X_0$. The total amount of material budget and each component's contribution to the tracker geometry are shown in Fig.~\ref{fig:EicC_material}.
With the current configuration, the material budget contributed by the barrel and disk staves is $< 5\%\,X_0$.  A potential future optimization to improve the performances of the barrel and end-cap tracker are ongoing.

In the simulation, the sensor silicon pixels have a pitch of $10\times10\, \rm \upmu m^2$ and silicon thickness of 50 $\rm \upmu m$. The pitch sizes of the readout of Micromegas are $150\, \rm \upmu m$ in both z and $r-\phi$ directions for the barrel region. Meanwhile, the Micromegas disk in the forward direction consists of many trapezoidal models, so the pitch size in $r-\phi$ direction ($50\,\rm \upmu m$) is much smaller than that in the $r$ direction (250 $\rm \upmu m$). 

The detector is mounted on a conical aluminum and carbon-fiber support structure with a thickness of 4.5 mm, which is tapered for $z > 58$ cm.  Although this support structure adds a huge amount of material to the detector as shown in Fig.~\ref{fig:EicC_material}, the projective design concentrates this material into a narrow pseudorapidity range at $|\eta|$ = 1.1. More realistic support structures and services are needed to be implemented in the future.

\subsection{Performance}

\subsubsection{Momentum resolution}

The performance studies were carried out within the EicCRoot software framework, which is being developed based on the FairRoot package~\cite{Al-Turany:2012zfk}. It is an object-oriented simulation, reconstruction, and data analysis framework. The geometry was implemented in GEANT4~\cite{GEANT4:2002zbu} and the charged particles (e.g. pions, electrons, protons, and muons) are generated from nominal IP and cover the entire detector acceptance by multiple generators provided by the EicCRoot. The momentum is set from 0 to 15 GeV/$c$ and distributed uniformly. The magnetic field is produced by a superconducting solenoid introduced by Section\ref{sec_B_field}. The interaction between the generated particles and the detector is handled by GEANT4, in which the multiple-scattering effect and energy loss of the track are taken into account. A hit is defined as the position where the particle enters the active area of the detector. Then, these hit positions are smeared according to the detector resolution. All the hits belonging to one track are selected according to the truth-track information and used for track fitting algorithm with the Kalman filtering method~\cite{Rauch:2014wta}.

\begin{figure}[H]
\centering
\includegraphics[width=0.45\textwidth]{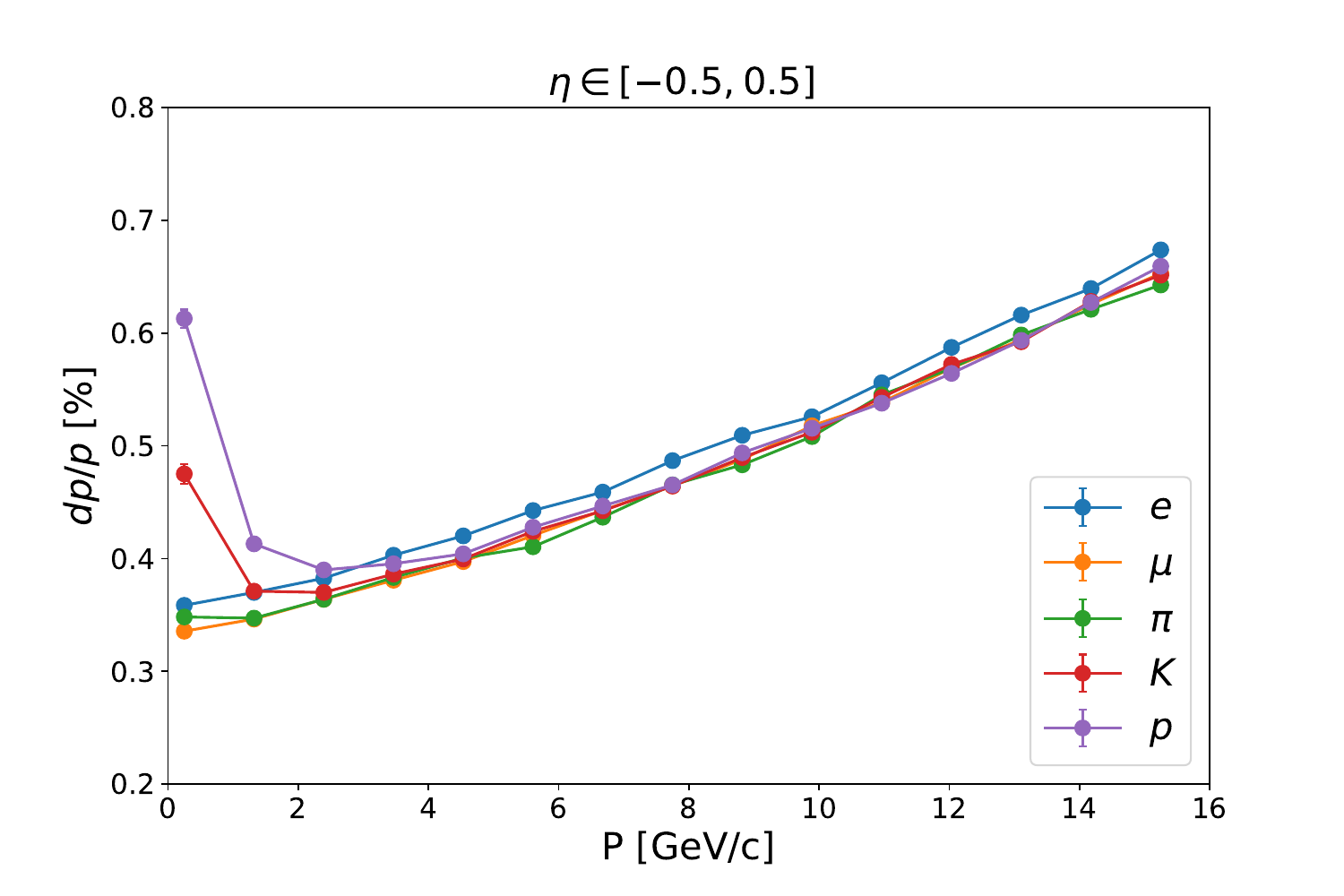}
\includegraphics[width=0.45\textwidth]{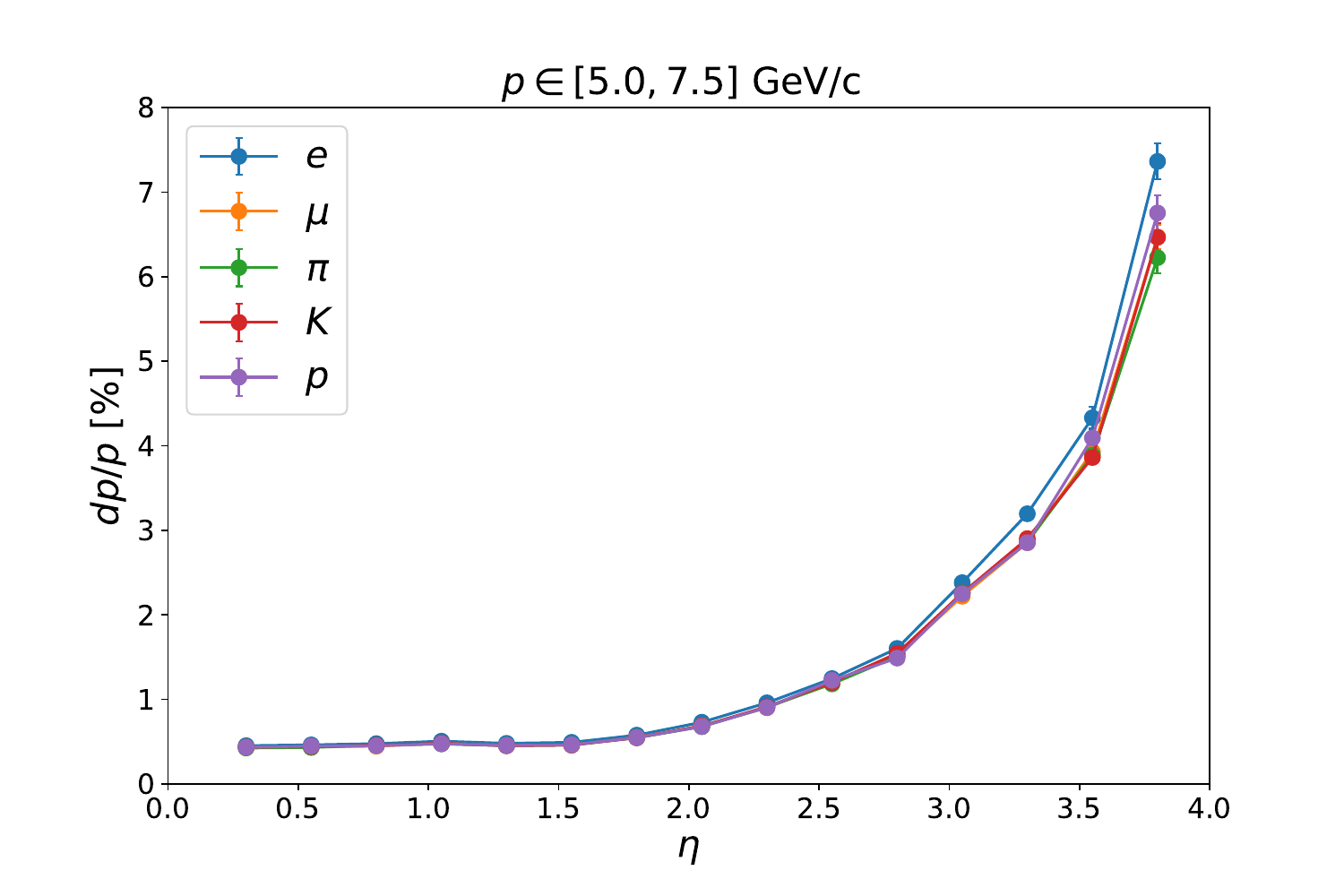}
\caption{Momentum resolution for different particles in the 1.5 T magnetic field. Left: $dp/p$ as a function of momentum in the $0 < \eta < 0.5$ range. Right: $dp/p$ as a function of pseudorapidity in the $5.0 < p < 7.5$ GeV/$c$ range.}
\label{fig:Mom_res}
\end{figure}

After the tracking fitting, the standard deviation of the $dp/p = (|\overrightarrow{p}_{truth}| - |\overrightarrow{p}_{reco}|) / |\overrightarrow{p}_{truth}|$ can be measured and defined as the momentum resolution, where $|\overrightarrow{p}_{truth}|$ and $|\overrightarrow{p}_{reco}|$ are the generated and reconstructed absolute value of the particle momentum, respectively. Figure ~\ref{fig:Mom_res} (top) shows the momentum resolution as a function of momentum for charged pions, electrons, muons, and protons in the pseudorapidity range $0.0 < |\eta| < 0.5$.  The multiple-scattering effect is more pronounced for protons and kaons below 2 GeV/$c$, therefore, worsening the resolution significantly. With the increase of the moment, the $dp/p$ rises almost linearly. This phenomenon is understandable because the measured sagitta will be decreased for the stiffer tracks (due to the higher momentum). Figure ~\ref{fig:Mom_res} (bottom) shows the momentum-resolution results as a function of pseudorapidity in the momentum range $5.0 < p < 7.5$ GeV/$c$. The momentum resolution is approximately constant when $\eta < 2$, and then it rises quickly. In addition, the electron momentum resolution is systematically worse than other particles in most of the studied range because of its smaller mass.  Overall, the performance is very similar for all kinds of particles.

\begin{figure*}[htbp]
\centering
\includegraphics[width=0.8\textwidth,trim={4cm 1.5cm 1.5cm 4cm}]{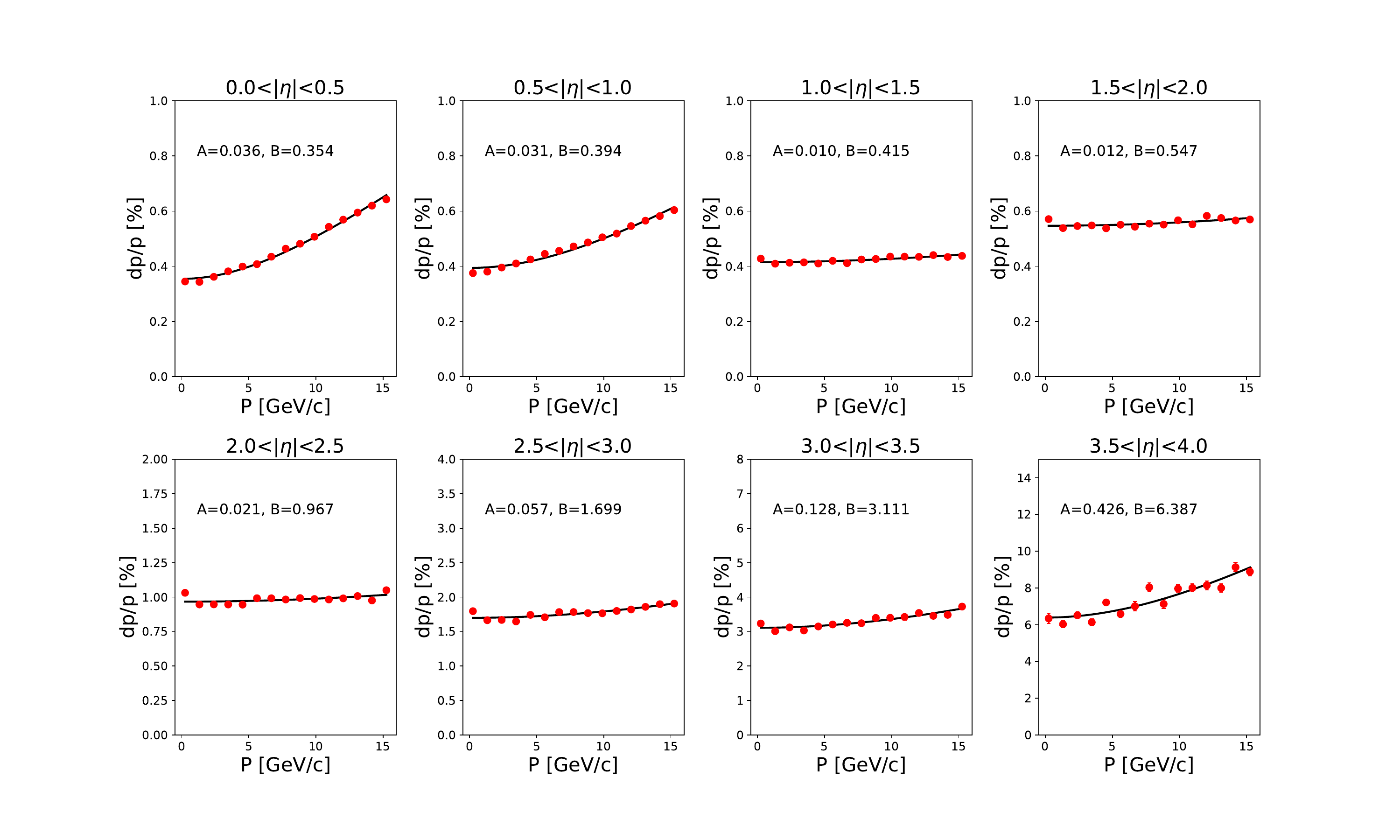}
\caption{Momentum resolution as a function of momentum in different pseudorapidity regions. The markers represent the resolutions obtained from the simulations, and the lines correspond to the best fit to the momentum-dependent resolution. The functional form used in the fits is $dp/p = Ap \oplus B$, and the parameters $A$ [\%/(GeV/$c$)] and $B$ [\%] are given in the plots.  }
\label{fig:res_p_vs_eta}
\end{figure*}

We also studied the momentum resolution in different $\eta$ regions for pions as shown in Fig.~\ref{fig:res_p_vs_eta}.  Momentum resolutions are typically parameterized by the function $dp/p = A\cdot p \oplus B$, where A and B are fit parameters and $\oplus$ indicates sum in quadrature. The fit results  are shown in the figure. In the case of the current magnetic field, the conceptual tracker design satisfies the physics requirements in the momentum range of $0 < p < 15$\,GeV/$c$ for $|\eta| < 2.5$.

\subsubsection{Spatial resolution and efficiency}

\begin{figure*}[htbp]
\centering
\includegraphics[width=0.8\textwidth,trim={4cm 1.5cm 1.5cm 4cm}]{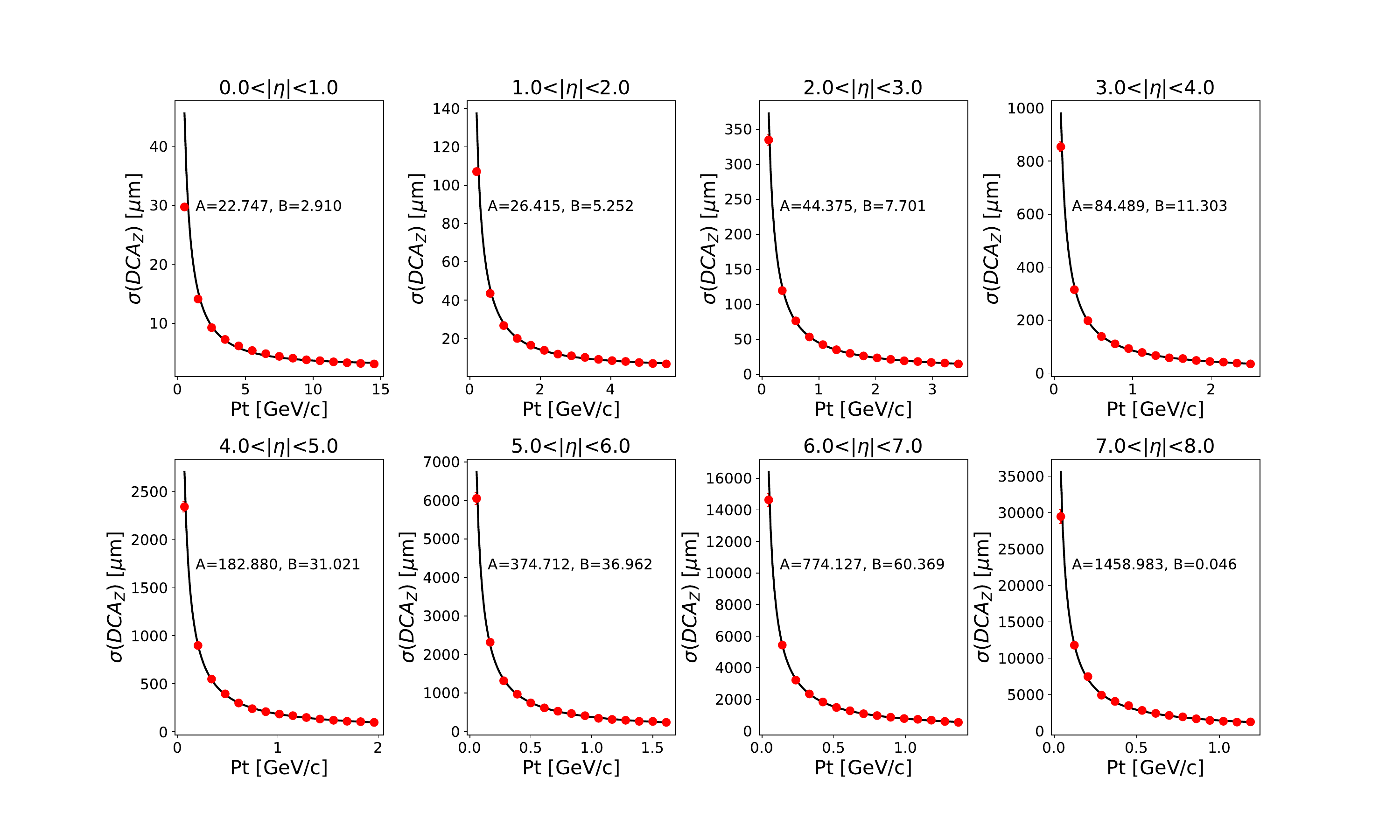}
\caption{Longitudinal DCA resolution as a function of transverse momentum in different pseudorapidity regions.  The markers represent the resolutions obtained from the simulations, and the lines correspond to the best fits to the momentum-dependent resolution. The functional form used in the fits is $\sigma(DCA) = A/p_T \oplus B$, and the parameters $A$ [$\rm \upmu m \cdot$ GeV/$c$] and $B$ [$\rm \upmu m$] are given in the plots. }
\label{fig:DCA_z}
\end{figure*}

\begin{figure*}[htbp]
\centering
\includegraphics[width=0.8\textwidth,trim={4cm 1.5cm 1.5cm 4cm}]{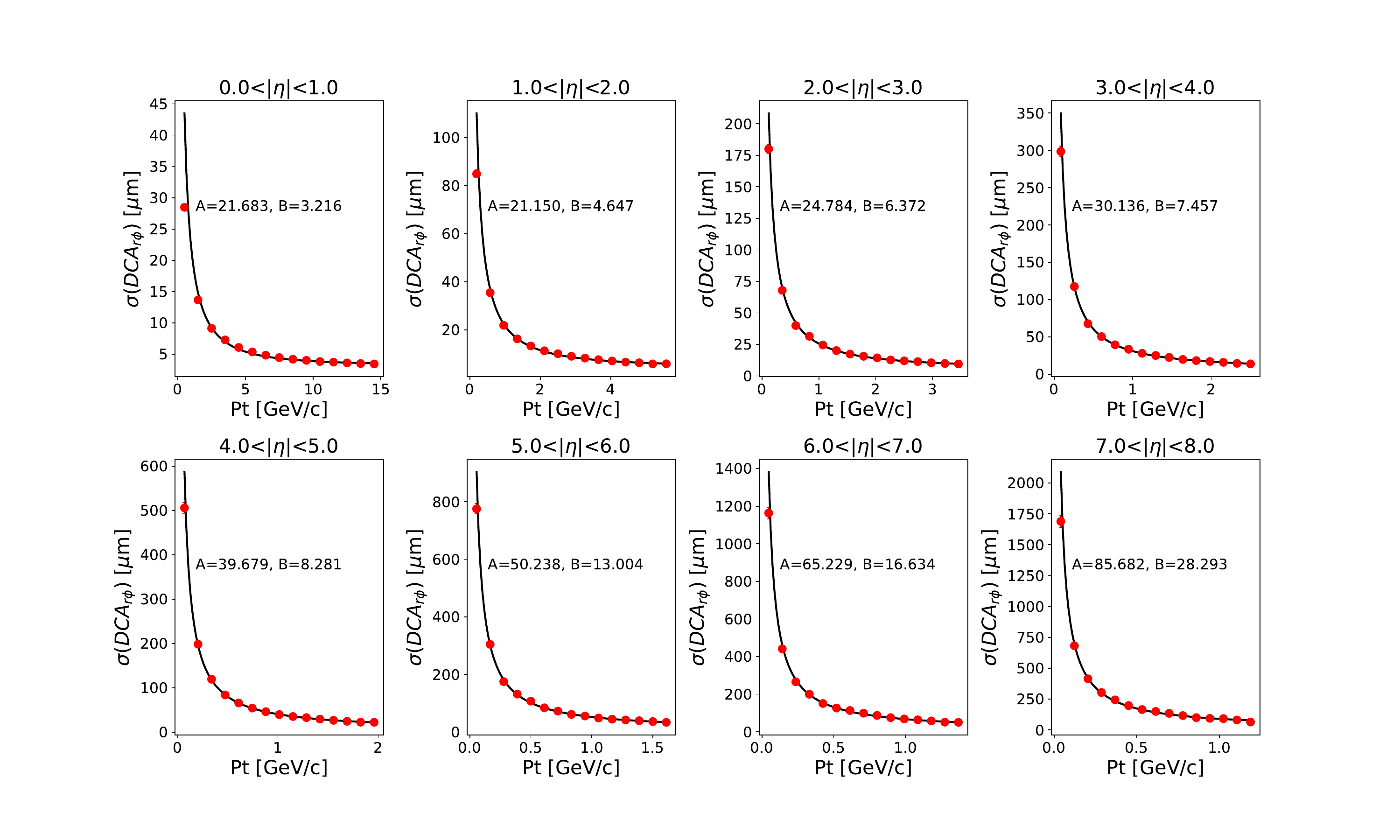}
\caption{Transverse DCA resolution as a function of transverse momentum in different pseudorapidity regions.  The markers represent the resolutions obtained from the simulations, and the lines correspond to the best fits to the momentum-dependent resolution. The functional form used in the fits is $\sigma(DCA) = A/p_T \oplus B$, and the parameters $A$ [$\rm \upmu m \cdot$ GeV/$c$] and $B$ [$\rm \upmu m$] are given in the plots. }
\label{fig:DCA_rp}
\end{figure*}

The tracking system is designed not only to measure the momentum of the charged particles but also to determine the primary vertex for an event as well as the secondary vertex for the long-life particle decay. This capability is crucial for background suppression and therefore improves the sensitivity significantly for the process with long-life particles. The spatial resolution can be studied by measuring the Distance of the Closest Approach (DCA), which is defined as the spatial separation between the primary vertex and the reconstructed track projected to the z-axis ($DCA_z$) or to the x-y plane ($DCA_{r\phi}$).
The DCA resolutions are determined as the standard deviation of normal functions that is used to fit to the $DCA_z$ and $DCA_{r\phi}$ distributions. DCA-resolution could be parameterized by the function $\sigma(DCA) = A/p_T \oplus B$, where $p_T$ stands for the transverse momentum. The results, including the fits and fitted parameters, for pions are shown in Figs.~\ref{fig:DCA_z} and \ref{fig:DCA_rp}.

\begin{figure*}[htbp]
\centering
\includegraphics[width=0.8\textwidth,trim={3cm 0.5cm 0.5cm 3cm}]{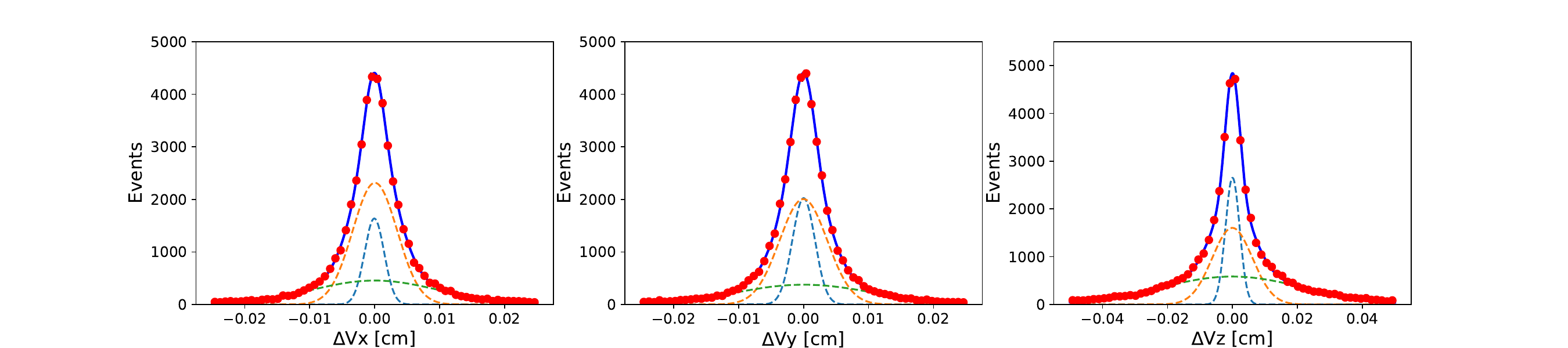}
\caption{Fit to the residuals of the primary vertex with three Gaussian functions. The markers correspond to the residual distribution from the simulation. The solid lines show the best fits to the residuals and the dashed lines are the three sub-components of the fit model.}
\label{fig:vertex}
\end{figure*}

\begin{figure}[htbp]
\centering
\includegraphics[width=0.4\textwidth]{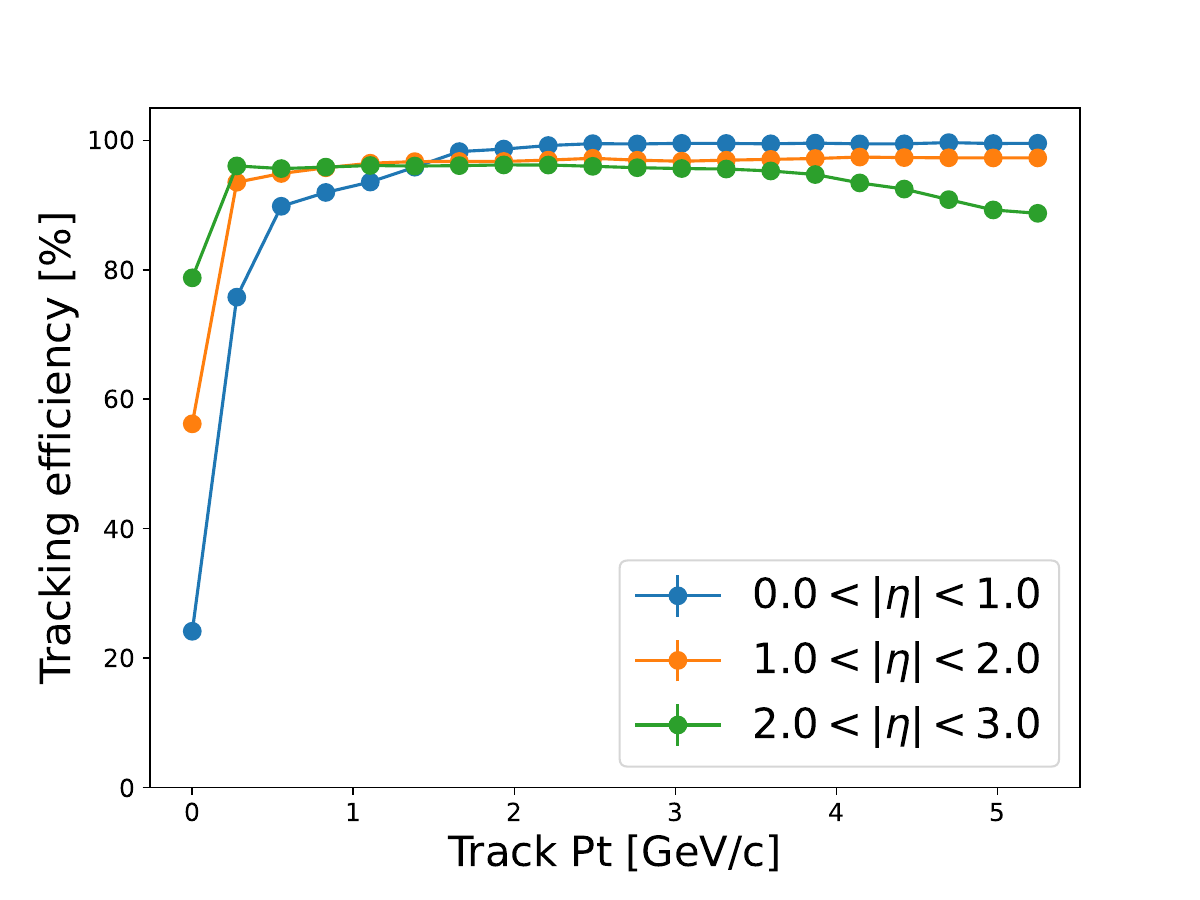}
\includegraphics[width=0.4\textwidth]{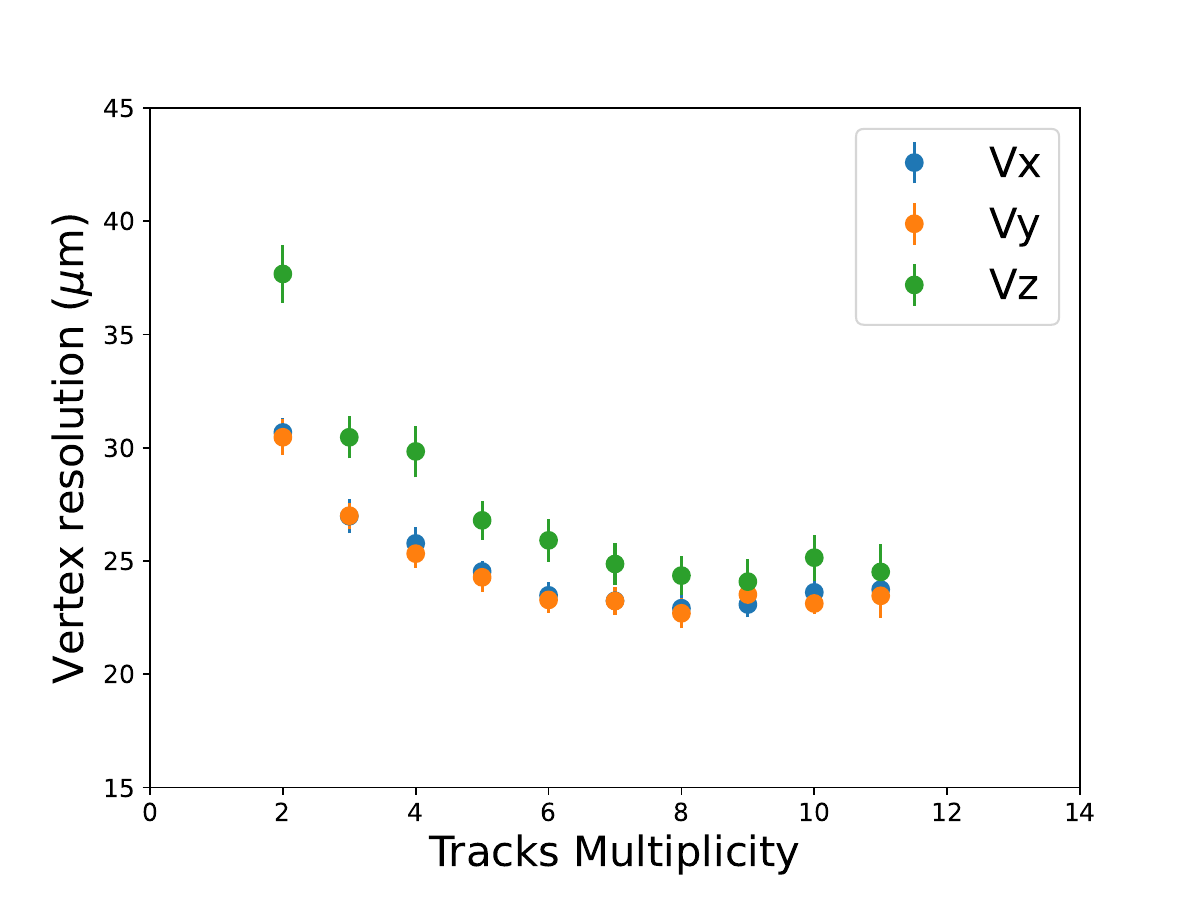}
\caption{Top: The tracking efficiency as a function of transverse momentum in several pseudo-rapidity bins determined in the full simulation with PYTHIA $e+p$ collision events at 18 GeV $\times$ 275 GeV with an event level selection of $Q^2 >1\, \rm{GeV^2}$. Bottom: the vertex resolution versus track multiplicity }
\label{fig:tracking_eff}
\end{figure}

The primary vertex represents the average vertex position of all the reconstructed charged tracks in an event assuming they originate from IP. To study the resolution of the primary vertex, millions of $e+p$ events at 3.5 GeV $\times$ 20 GeV collisions are generated by the PYTHIA generator in the EicCRoot framework. In the simulation, we require $Q^2 >1\,\rm{GeV^2}$, and all the hadrons and scattering electrons are used to reconstruct the primary vertex. The primary vertex is obtained by fitting all the tracks with the Kalman filtering method~\cite{Waltenberger:2011zz}. The vertex residual distribution with respect to the generated truth vertex for track multiplicity of three is shown in Fig.~\ref{fig:vertex}. Since these distributions contain non-Gaussian tails, it is difficult to be parameterized as a normal distribution. Therefore, three coherent Gaussian functions are used to parameterize these residuals. To investigate the resolution change with respect to the track multiplicity, we fit the peak of these distributions with a single Gaussian function and define the standard deviation as the primary vertex resolution.
The bottom plot of Fig.~\ref{fig:tracking_eff} shows the resulting primary vertex resolution as a function of track multiplicity in events. 
The primary vertex resolution gets improved with the increase of the tracking number and the values vary from $\rm 35\, \upmu m$ to $\rm 25\,\upmu m$ in the studied range. In the fast simulation, the primary vertex is smeared by sampling the fitted three Gaussian functions.

Figure~\ref{fig:tracking_eff} top plot shows the charged pion tracking efficiency in different $\eta$ regions. It should be noticed that pattern recognition efficiency is not included in this study. The efficiency loss is mainly due to the detector acceptance and the track fitting quality. To reconstruct a track, we require that at least 3 hits should be found. Therefore, in the central part of the detector ($|\eta|<$1), the efficiency in low $p_T$ is worse than in the forward/backward region. It is related to the minimum $p_T$ threshold needed for a track to reach the outer layer.
For the $|\eta|>1$ region, the loss of efficiency at higher $p_T$ because of the lower acceptance at the edge of the barrel-to-endcap transition region and also at the small radii where the beam pipe openings in the silicon disk. These tracking efficiencies obtained from this simulation were applied in the following performance studies through fast simulation.

\section{Impact study on gluon distributions} \label{sec:physics}

\begin{figure}[htbp]
\centering
\includegraphics[width=0.35\textwidth]{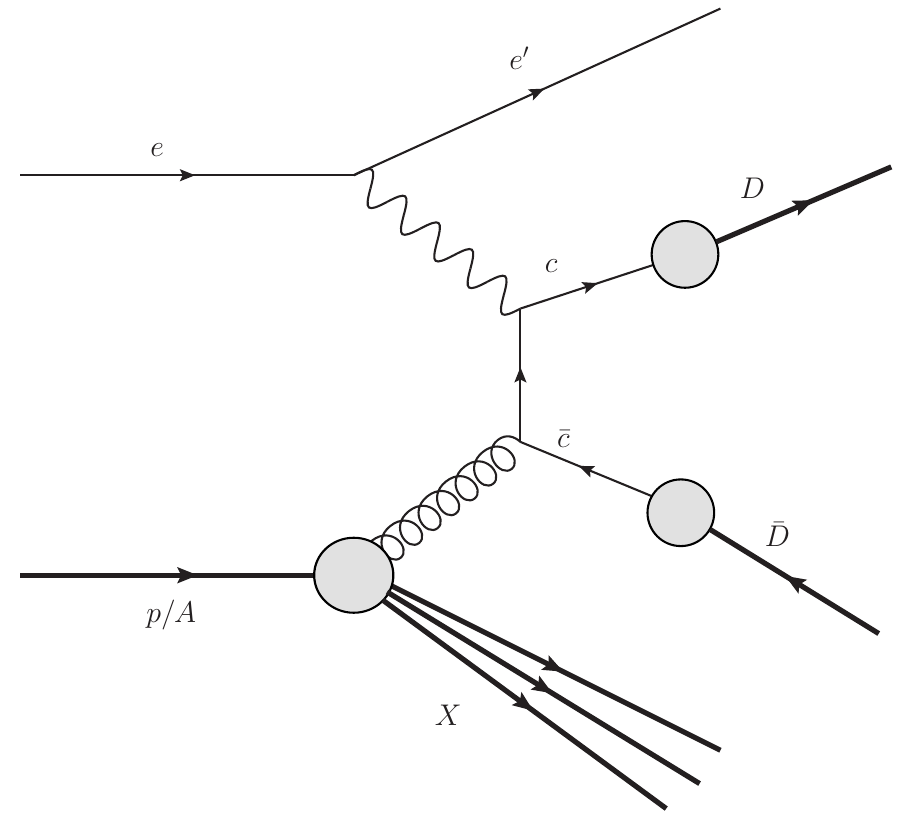}
\caption{The leading-order diagram for charm and anticharm pair production in $e+p/A$ deep inelastic scatterings.  }
\label{fig:PGF}
\end{figure}

In the deep inelastic scattering, charm or anti-charm quarks can be produced through PGF process at the leading order: $\gamma^*+g \rightarrow c+\bar{c}$. The charm or anti-charm quarks are further fragmented into the D mesons. The Feynman diagram of PGF is shown in Fig.~\ref{fig:PGF}.  By tagging a D meson in the final state for which tracking is critical, one can access unpolarized/polarized gluon distributions in the nucleon, depending on different experimental observables. In this section, we will discuss the impact studies on unpolarized gluon distribution via $F_2^{c\bar{c}}$ measurements and the $\Delta g$ distribution via $A_1^{c}$  measurements at the EicC.

Without conducting entirely new Parton Distribution Function (PDF) fits, we employ well-established re-weighting techniques~\cite{Giele:1998gw, Ball:2010gb, Ball:2011gg} to facilitate our studies. The central values and uncertainty bands of the distributions can be determined by statistically averaging and calculating the variance of a replica set. This set can be generated through random smearing of the data points within their error bands, followed by re-fitting the data. Re-weighting procedures leverage Bayesian interference to incorporate the information from a new data set into the probability distribution of the initial replica set. This is achieved by assigning appropriate weights to different replicas. The resulting re-weighted replica set, which incorporates the new data information, is then utilized to compute the central values and uncertainties of the updated distributions while preserving the statistical rigor of the original set. However, it should be noted that if a large number of replicas are suppressed due to vanishing weights, the statistical relevance of the new replica set is compromised, necessitating a new fit.

\begin{figure}[htb]
\centering
\includegraphics[width=0.35\textwidth]{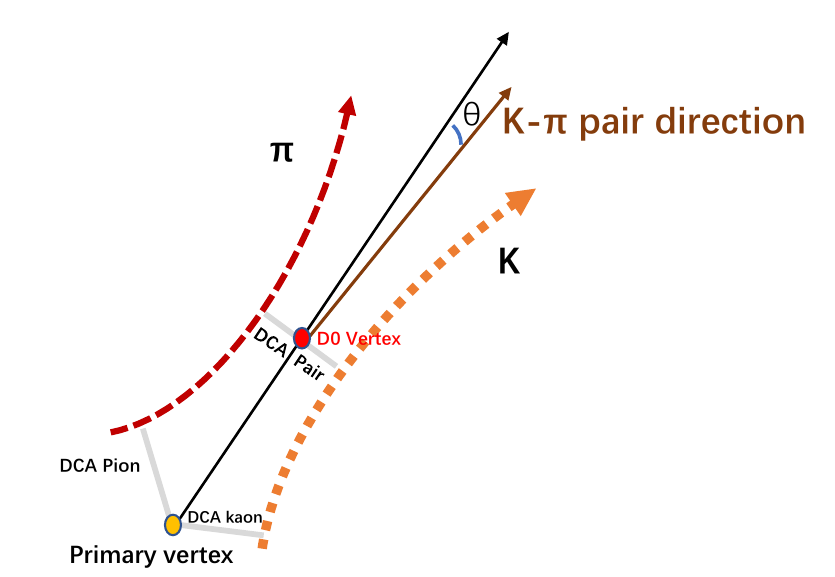}
\caption{Illustration of the $D^0$ to $K\pi$ two-body decay topology.}
\label{fig:D0_illustration}
\end{figure}

\begin{figure}[htb]
\begin{center}
\centering
\includegraphics[width=0.35\textwidth]{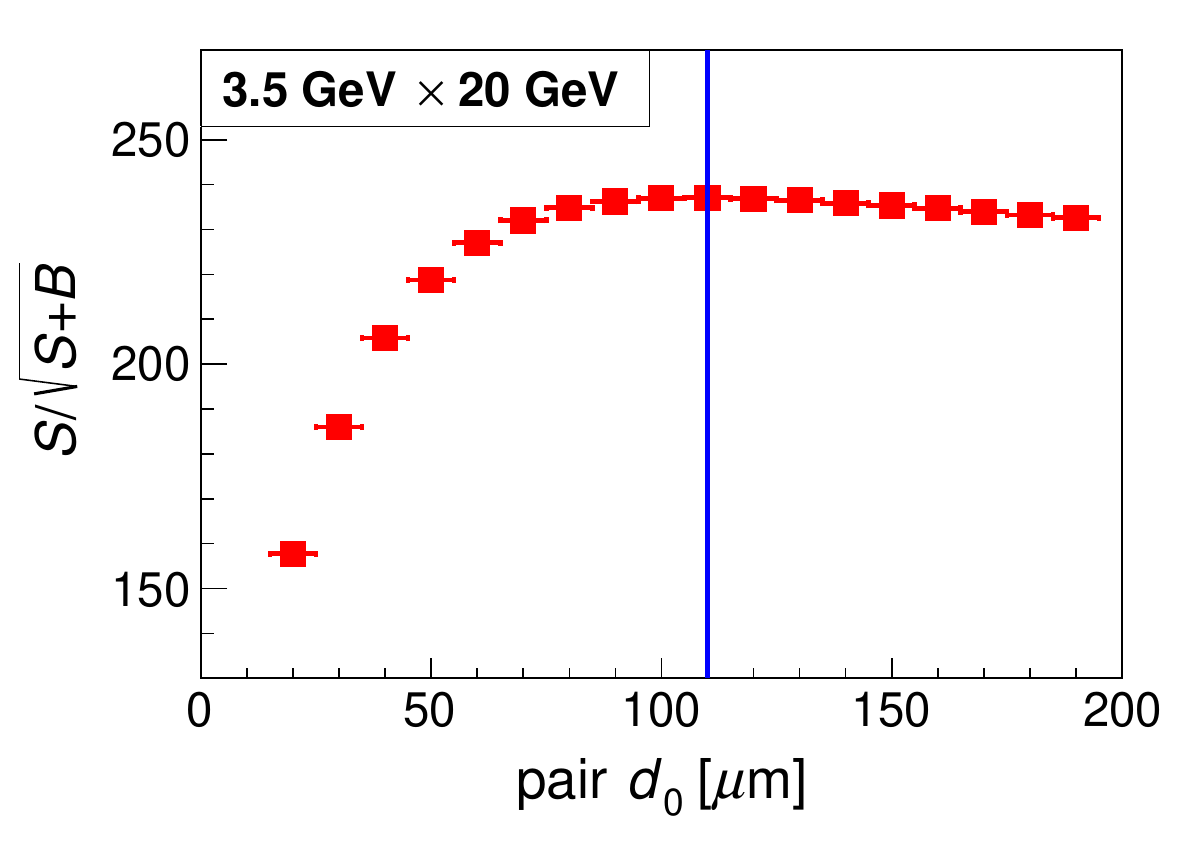}
\includegraphics[width=0.35\textwidth]{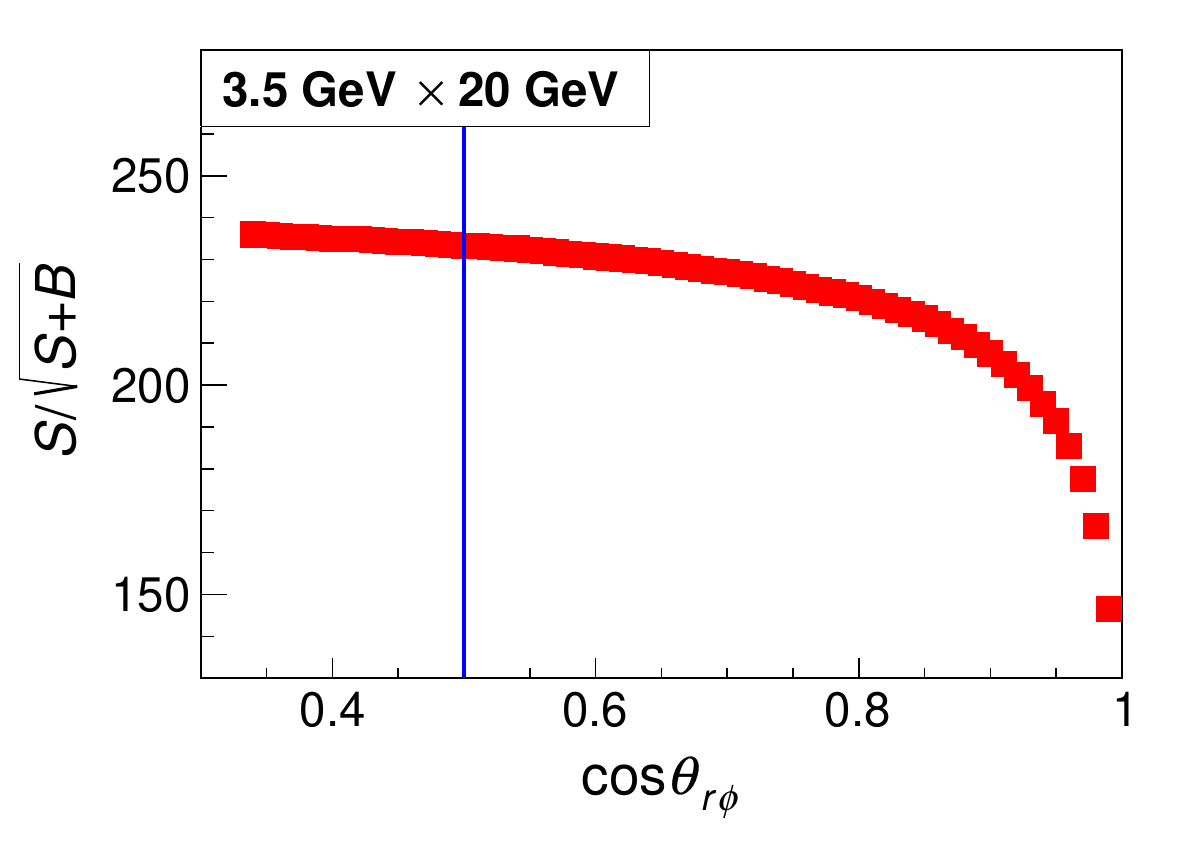}
\includegraphics[width=0.35\textwidth]{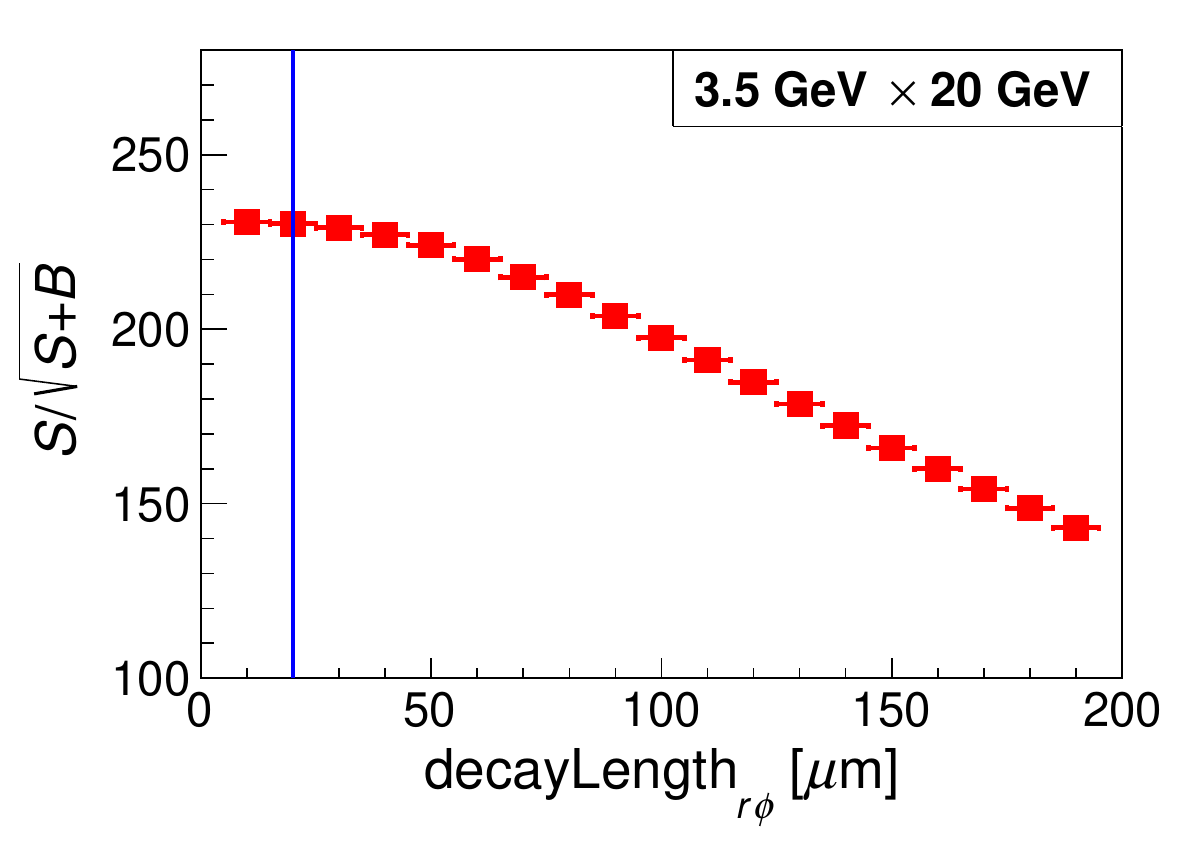}
\caption{The figure-of-merit, FOM = $S/\sqrt{S+B}$, distributions in the criteria optimization for $K\pi$ pair-DCA (top), the cos$\theta_{r\phi}$ (middle), and $D^0$ Decay-Length$_{r\phi}$ in the transverse plane (bottom). The blue lines show the position at the maximum FOM.}  
\label{fig_cut_1.5T_3.5_20}
\end{center}
\end{figure}

\begin{figure}[htb]
\begin{center}
\centering
\includegraphics[width=0.35\textwidth]{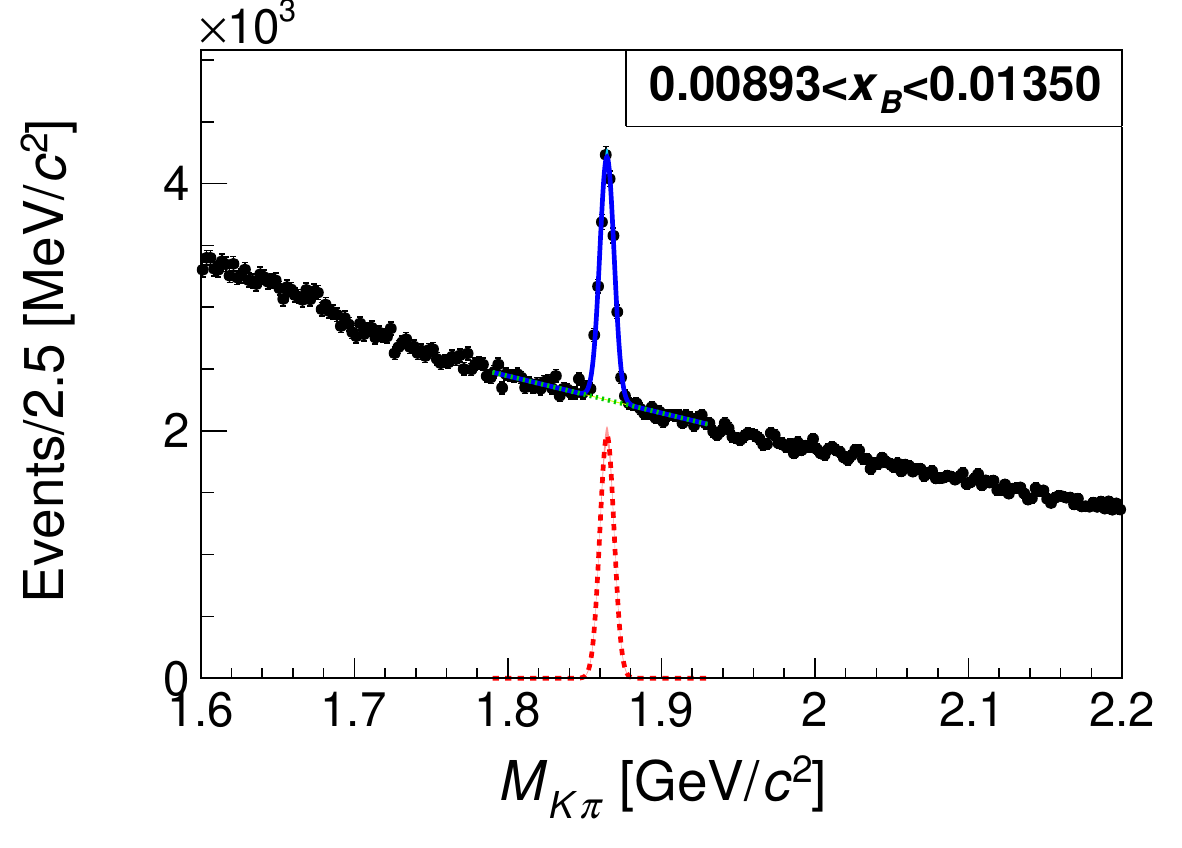}
\includegraphics[width=0.35\textwidth]{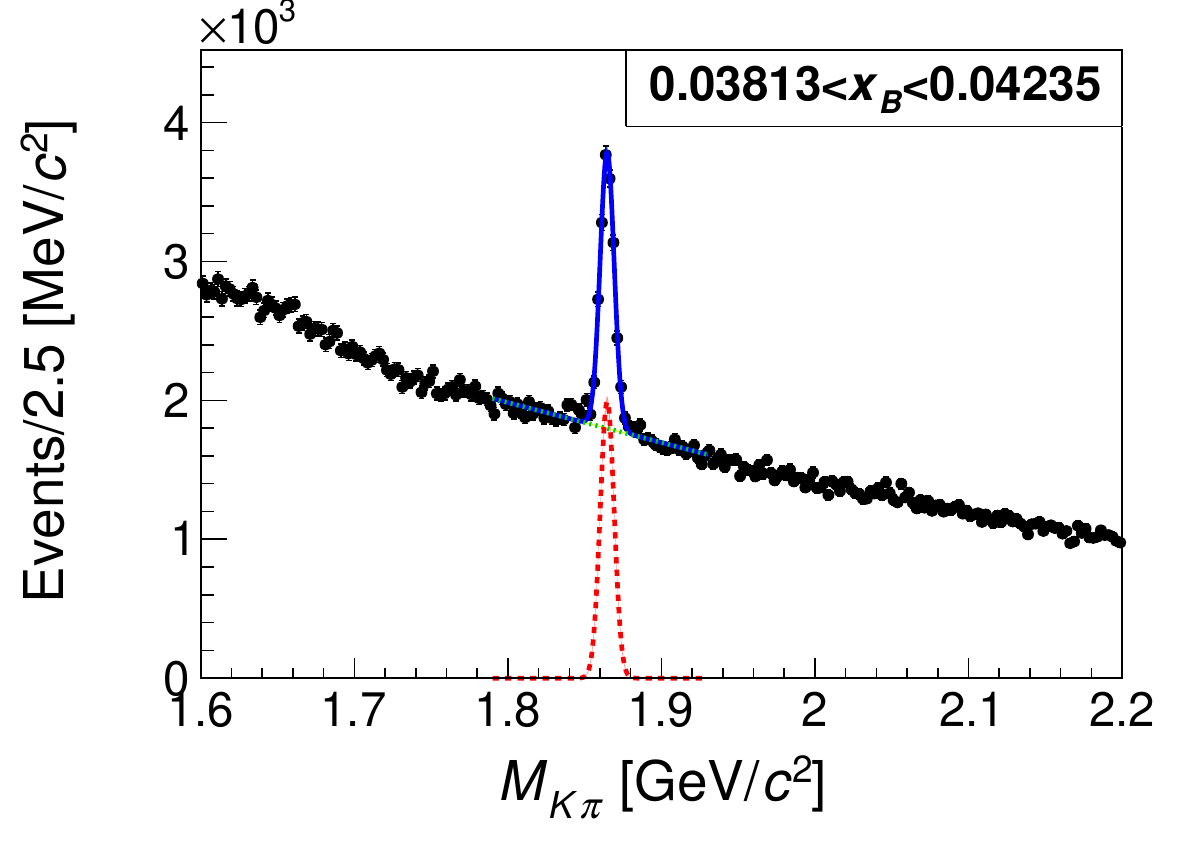}
\includegraphics[width=0.35\textwidth]{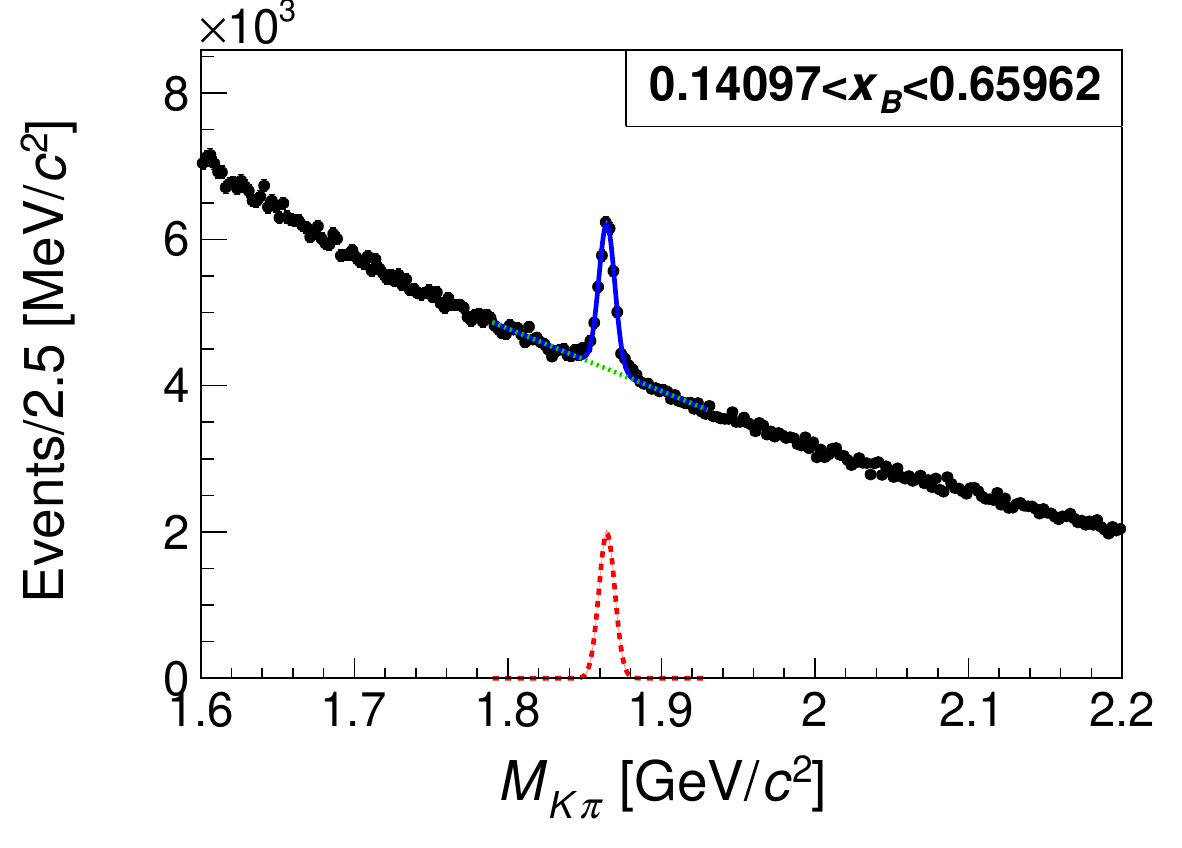}
\caption{Fits to the $K\pi$ invariant-mass distributions in a few different Bjorken-$x$ bins for 3.5 GeV $\times$ 20 GeV $e$+$p$ collisions. The red and green dashed curves are the signal (Gaussian) and background (linear) fits, and the blue curve is the sum.  }  
\label{fig:fit_3.5_20}
\end{center}
\end{figure}

The EicC pseudo data for reweighting analysis was produced as follows:
The geometry of the tracking system has been implemented in the GEANT4 and studied within the full
Monte-Carlo framework for detector simulation, as discussed in Section \ref{sec:tracking}. The full simulation yields the
detector response for momentum resolution, single-track DCA resolution, primary vertex resolution, as
well as tracking efficiency. Afterward, the implementation of resolution parametrizations in a fast-smearing simulation framework enables the generation of ample statistics, facilitating comprehensive studies for physics projections.

In our study, two beam energy configurations have been used for electron-proton collisions: 3.5 GeV $\times$ 20 GeV and
5 GeV $\times$ 25 GeV. For the nuclear gluon distribution study, we use electron-gold
collisons with 3.5 GeV $\times$ 10.35 GeV/u and 3.5 GeV $\times$ 12.93 GeV/u energy configurations.
The data was generated by pythiaeRHIC ~\cite{pythiaerhic} (PYTHIA V6.4) and then fed into the fast-smearing framework to accommodate detector
response. We take advantage of the $K\pi$ two-body-decay to identify the $D^0$($\bar{D^0}$). 
The decay-topology is illustrated in Fig. \ref{fig:D0_illustration}.
To obtain a data sample with a good signal-to-background ratio, three decay topological distributions were investigated: the $K\pi$ pair-DCA, the $D^0$ Decay-Length ${r\phi}$ in the transverse plane, and the angle $\cos(\theta_{r\phi})$, where $\theta$ represents the angle between the $D^0$ direction with respect to the primary vertex and the momentum vector of the $K\pi$ pair. In practice, taking 3.5 GeV $\times$ 20 GeV electron-proton collision as an example, the criteria of these topological requirements are optimized by maximizing the Figure-Of-Merit, FOM = $S/\sqrt{S+B}$, where $S$ is the number of $D^0$ and B is the estimated background from the simulation. The results of optimization are shown in Fig.~\ref{fig_cut_1.5T_3.5_20}.

In addition to the $D^0$-decay topology requirements, the following kinematic constraints were applied during the analysis: $Q^2>$ 2 GeV$^2$, 0.05 $<$ y $<$ 0.8
and $W^2 >$ 4 GeV$^2$. The pion/kaon separation was assumed to be feasible up to the momentum limits: 4, 6, 15 GeV/$c$ in pseudo-rapidity regions (-3,-1), (-1,1) and (1,3) respectively. Here, the pseudo-rapidity is positive in the ion-going direction.
Following the selection requirements, the data were binned in Bjorken-$x$ and $Q^2$. Within each bin, the $K\pi$ invariant mass spectrum was fitted to a Gaussian function for the signal, along with a linear background, as shown in Fig.~\ref{fig:fit_3.5_20}, which depicts electron-proton collisions with an energy of 3.5 GeV $\times$ 20 GeV for various Bjorken-$x$ bins.

\subsection{Unpolarized proton gluon distribution function}

Structure functions provide valuable insights into the partonic structure of hadrons. The structure function $F_2(x,Q^2)$ specifically characterizes the interaction between transversely polarized photons and the proton.
The open-charm contribution, $F_2^{c\bar{c}}$ can be defined in terms of the inclusive double differential $c\bar{c}$ cross-section in $x$ and $Q^2$ by
\begin{widetext}
\begin{equation}
\frac{d^2\sigma^{c\bar{c}}}{dx\,dQ^2}=\frac{2\pi\alpha^2}{xQ^4} \left\{[1+(1-y)^2]F_2^{c\bar{c}}(x,Q^2)-y^2F_L^{c\bar{c}}(x,Q^2)\right\} .
\end{equation}
\end{widetext}

The equation can be rewritten to form a reduced cross-section as

\begin{equation}
\begin{aligned}
\sigma_r^{c\bar{c}}(x,Q^2)&=\frac{d^2\sigma^{c\bar{c}}}{dx\,dQ^2} \times\frac{xQ^4}{2\pi\alpha^2[1+(1-y)^2]}\\
&= F_2^{c\bar{c}}(x,Q^2)-\frac{y^2}{1+(1-y)^2}F_L^{c\bar{c}}(x,Q^2)
\end{aligned}
\end{equation}

To extract $F_2^{c\bar{c}}$ at a fixed $Q^2$ and $x$, the Rosenbluth technique~\cite{PhysRev.79.615} requires a minimum of two collision energies. In this study, beam configurations of 3.5 GeV $\times$ 20 GeV and 5 GeV $\times$25 GeV were chosen, each with an integrated luminosity of 10 fb$^{-1}$. 
The differential cross section can be determined experimentally

\begin{equation}
\frac{d^2\sigma^{c\bar{c}}(x,Q^2)}{dxdQ^2}=\frac{dN(D^0+\bar{D}^0)/2}{L\cdot \varepsilon\cdot  B(D^0\rightarrow K\pi) \cdot f(c \rightarrow D^0) \cdot dxdQ^2}.
\end{equation}

Here, $L$ is the integrated luminosity, $\varepsilon$ is the detection efficiency, $B(D^0\rightarrow K\pi)$ is the branching fraction of $D^0$ to $ K\pi$ from PDG~\cite{10.1093/ptep/ptac097}, and $f(c \rightarrow D^0)$ is the $D^0$ fragmentation fraction in PYTHIA~\cite{Sj_strand_2006}. The $dN(D^0+\bar{D}^0)$ in the numerator is the number of detected D mesons in the specific $(x,Q^2)$ bin which is determined from a fit to the $K\pi$ invariant mass spectrum.

\begin{figure*}[htb]
\centering
\includegraphics[width=0.3\textwidth,trim={0.2cm 0.5cm 0.5cm 0.2cm}]{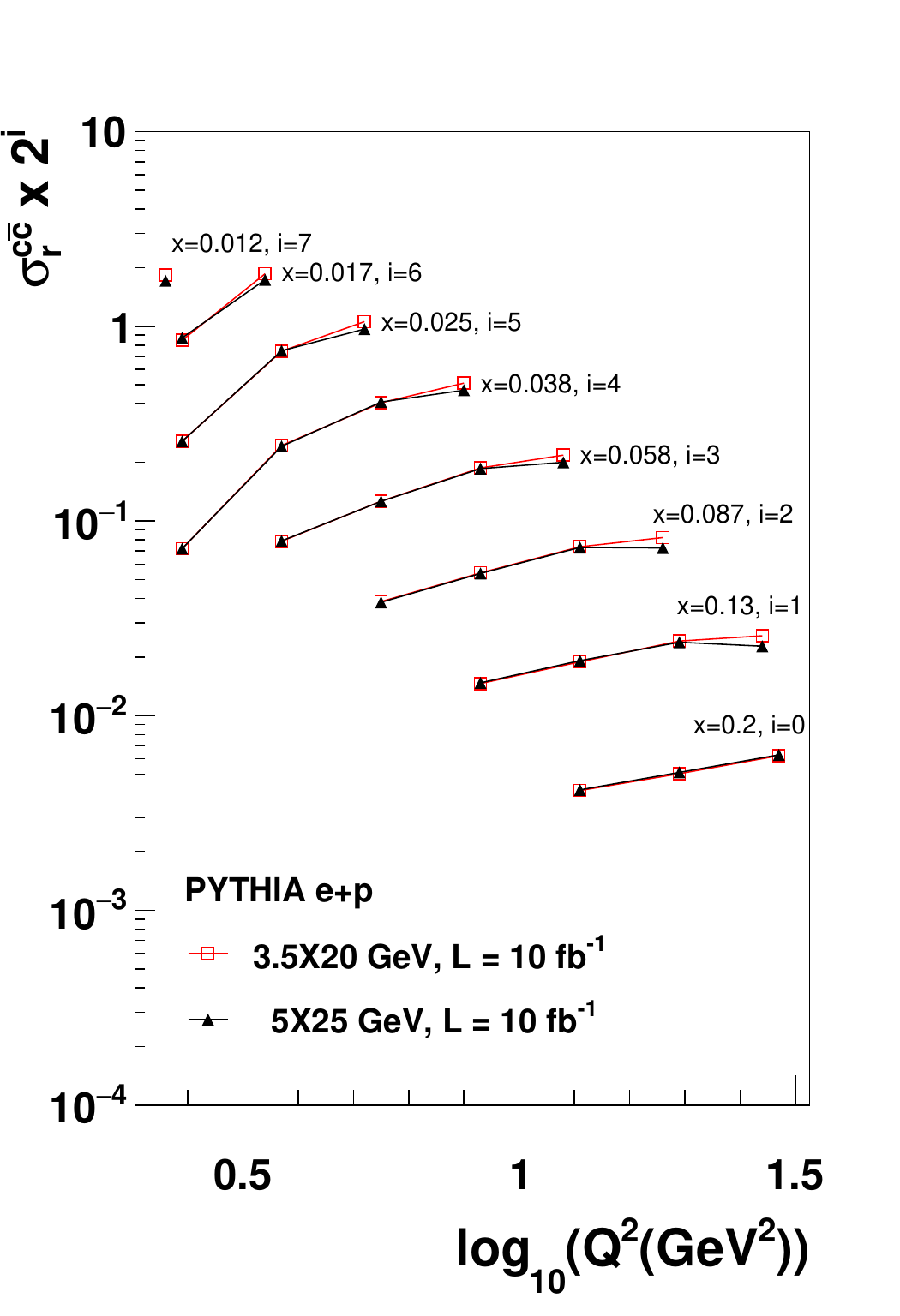}
\includegraphics[width=0.3\textwidth,trim={0.2cm 0.5cm 0.5cm 0.2cm}]{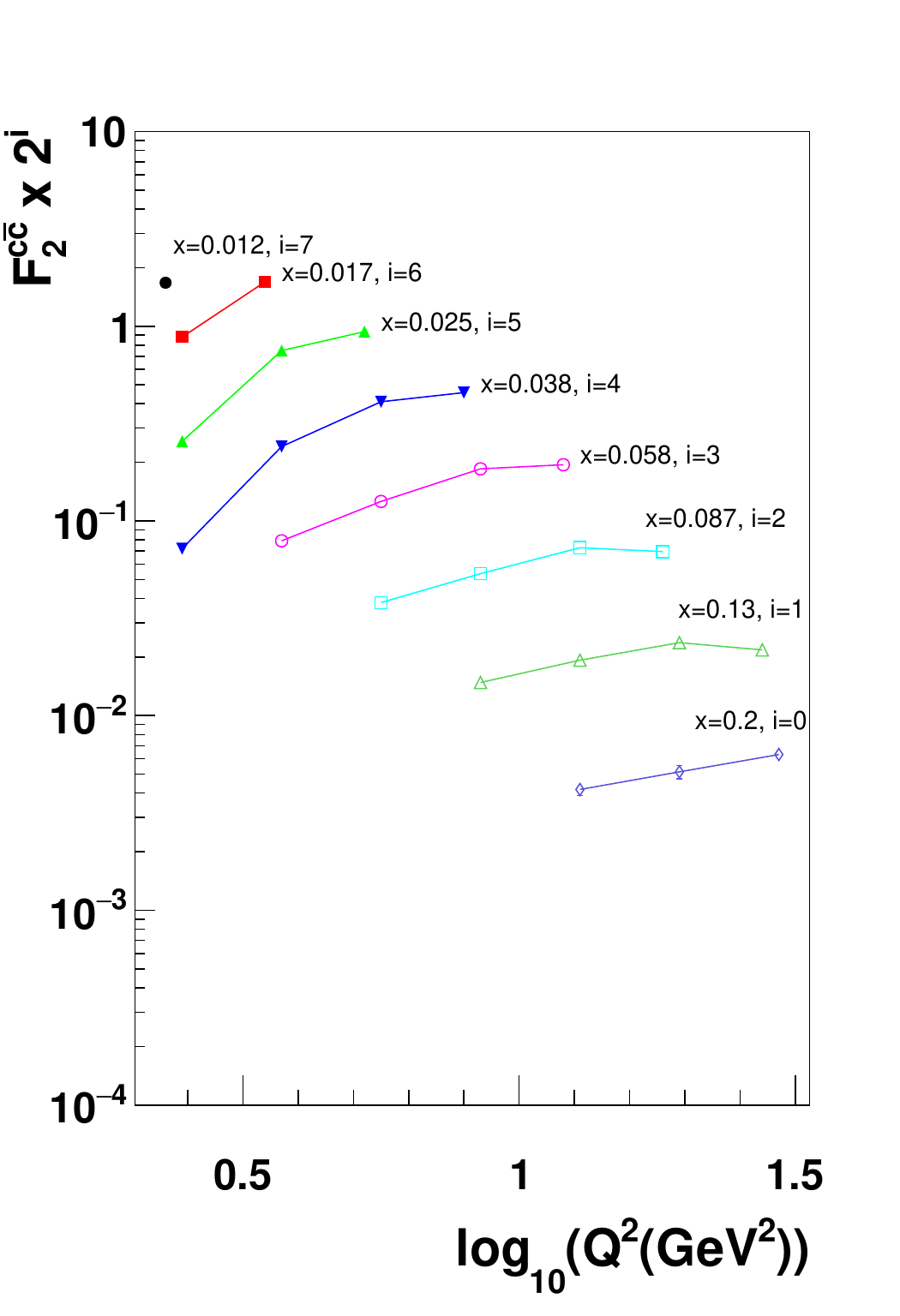}
\caption{The reduced cross sections (left panel) and the projected $F_{2}^{c\bar{c}}$ (right panel) for the proton beams as a function of the photon virtuality. We show two beam configurations for the reduced cross sections from which we then extract the structure function.}
\label{fig:F2ccbar_ep}
\end{figure*}

In Fig.~\ref{fig:F2ccbar_ep}, we present the projected measurements based on one year’s worth of data collected at the EicC. The left plot shows the reduced cross sections for each $Q^2$ and $x$ bins at the two energies. The red boxes indicate the low energy pseudo-data, while the black triangles are the high energy counterpart. Using the reduced cross sections at these two energies, the $F_{2}^{c\bar{c}}$ for the proton is extracted and shown in the right plot of Fig.~\ref{fig:F2ccbar_ep}.

By utilizing these measurements, we can enhance our understanding of the unpolarized proton gluon PDF, denoted as $g(x,Q^2)$, through reweighting techniques. For this initial estimate, our focus is on evaluating the impact on the most recent CTEQ PDF set~\cite{Hou:2019efy}. We employ the \texttt{mcgen} program~\cite{Gao:2013bia, Hou:2016sho} to convert the Hessian PDF representation into a Monte Carlo replica representation. Furthermore, we provide theoretical predictions, accurate to next-to-leading order (NLO) precision, for all the required observables using \texttt{yadism}~\cite{candido_alessandro_2022_6285149}.

\begin{figure*}[htb]
\centering
\includegraphics[width=0.8\textwidth,trim={2cm 0 0 2cm},clip]{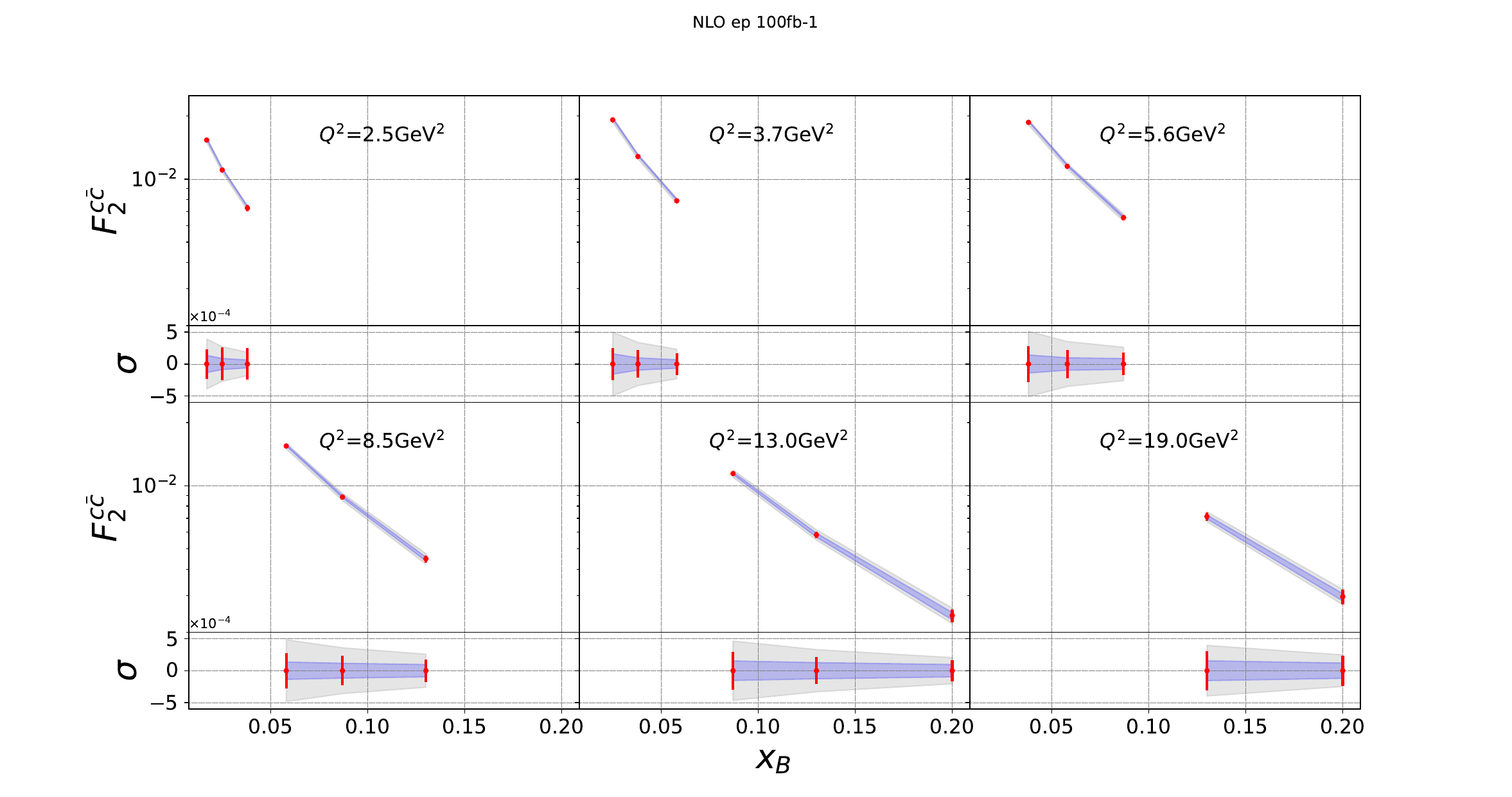}
\caption{The predictions for the projected measurement of the structure-function: in red are the projected experimental uncertainties, in light grey (blue) is the theoretical predictions using the \texttt{CT18ANLO}~\cite{Hou:2019efy} PDF set before (after) re-weighting.}
\label{fig:ep_F2_CT18ANLO}
\end{figure*}

In Fig.~\ref{fig:ep_F2_CT18ANLO}, we present a comparison between the projected experimental uncertainty for each bin and the corresponding PDF uncertainty before and after reweighting. The subplots are organized based on a fixed value of virtuality $Q^2$, while the horizontal axis represents the variation of Bjorken-$x$. Here, our kinematics are constrained within the region $0.01 < x < 0.2$ due to the specific beam configurations. Moreover, it should be noted that PDF uncertainties are influenced by assumptions inherited from the fitting procedure.

\begin{figure}[htb]
\centering
\includegraphics[width=0.45\textwidth,trim={0 0 0 2cm},clip]{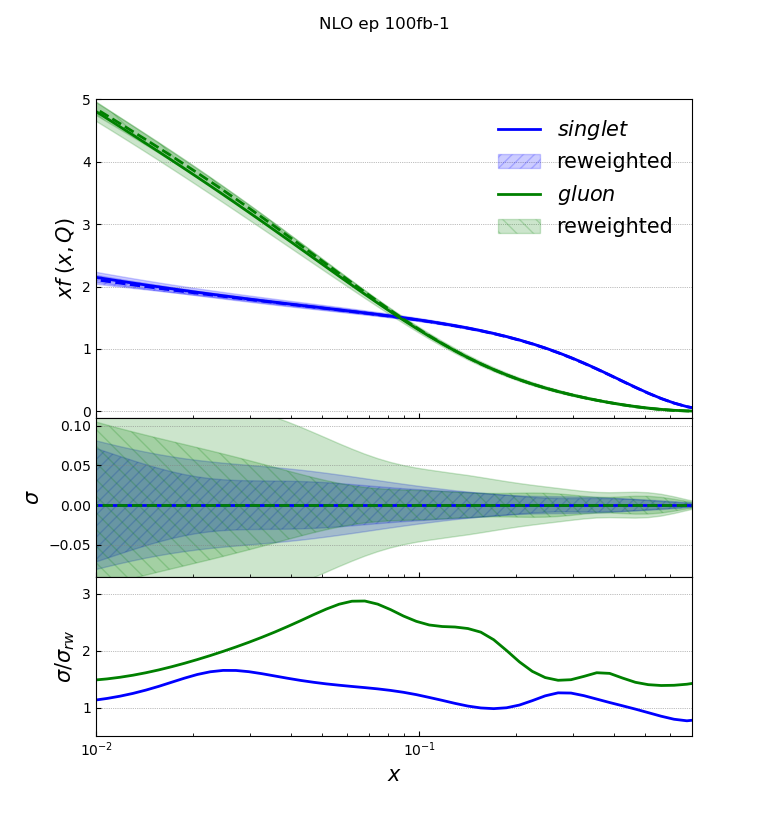}
\caption{Impact on CTEQ~\cite{Hou:2019efy} singlet $\Sigma$ and gluon $g$ PDFs using a NLO computation. The hashed bands show the impact of the pseudo data on the distributions' uncertainties, whereas the solid bands show the original uncertainty. In the middle plot, we show the absolute uncertainty $\sigma$. The bottom plots show the ratios between the absolute uncertainties before and after reweighting.}
\label{fig:CTEQNLO}
\end{figure}

Figure~\ref{fig:ep_F2_CT18ANLO} allows us to extract the influence of the measurement on the underlying PDF, as demonstrated in Fig.\ref{fig:CTEQNLO}. The upper plots depict the singlet and gluon PDF of the CTEQ PDF set~\cite{Hou:2019efy}, while the middle and lower plots focus on the corresponding uncertainties. Through the reweighting procedure, the number of replicas contributing to the uncertainties are reduced from 100 to approximately effective 25. Consequently, we observe a modest impact on the singlet PDF, resulting in an uncertainty reduction of around 50\%. In contrast, the gluon PDF shows a significant impact, with a maximum uncertainty reduction by a factor of 3. It is important to note that despite the use of a general-mass variable flavor number scheme (GM-VFNS)~\cite{Forte:2010ta} in computing the theory predictions, the relatively small virtuality $Q^2$ still leads to the dominance of the PGF mechanism, directly influencing the gluon PDF.

\subsection{Unpolarized nuclear gluon PDF}

\begin{figure*}[htb]
\centering
\includegraphics[width=0.7\textwidth,trim={1cm 0 0 1cm}]{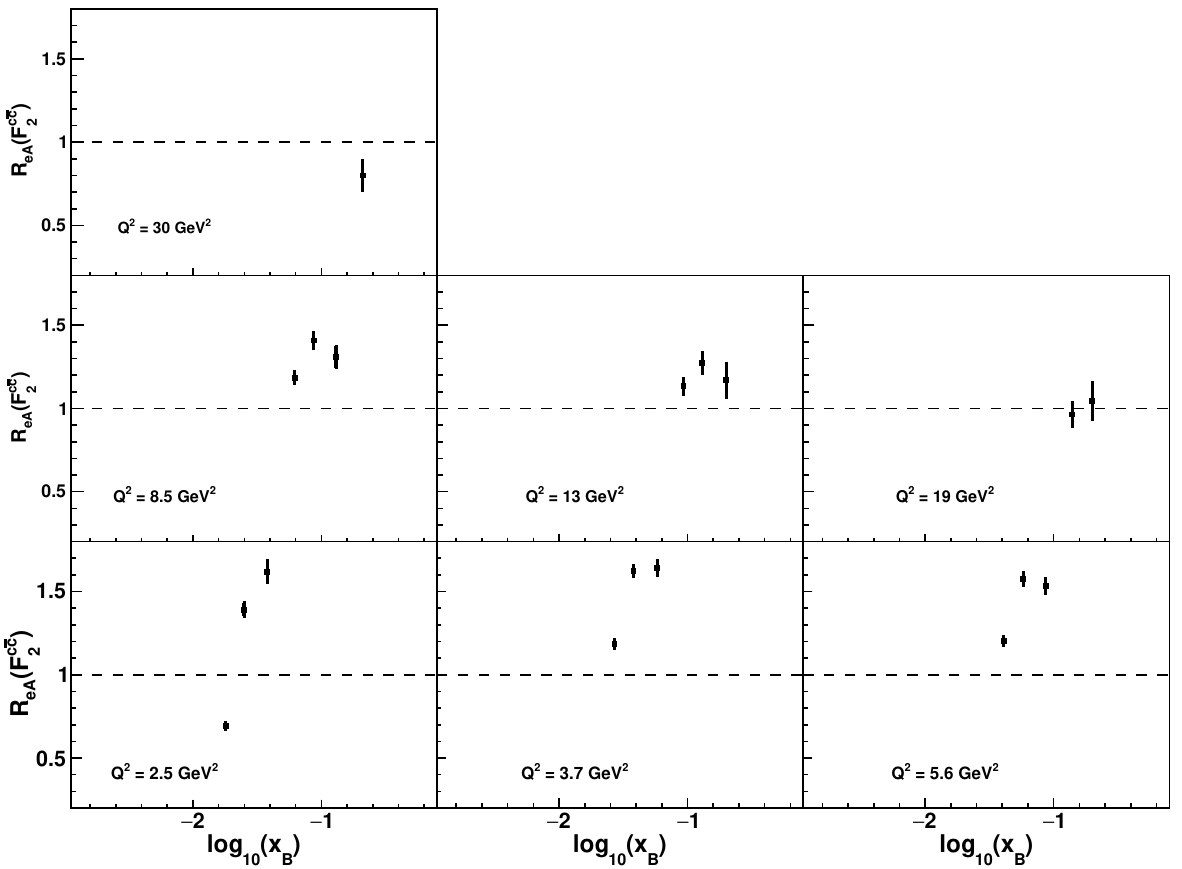}
\caption{Ratio of $F_{2}^{c\bar{c}}$ for the nuclear case (gold) towards the proton case.}
\label{fig:ReA_F2ccbar}
\end{figure*}

Likewise, by performing measurements on reduced cross sections in electron-ion collisions, it is possible to extract the nuclear $F_{2}^{c\bar{c}}$, thereby gaining insights into the unpolarized nuclear gluon distributions. In this context, we demonstrate the potential of the EicC through electron-gold collisions. For this purpose, we generate two sets of pseudo-data for e+Au collisions: 3.5\,GeV$\times$10.35 GeV/u and 5\,GeV$\times$12.93 GeV/u, each with an integrated luminosity of 10 fb$^{-1}$. The resulting projected ratios of $F_{2}^{c\bar{c}}$ in e+Au collisions relative to e+p collisions are displayed in Fig.~\ref{fig:ReA_F2ccbar}.

\begin{figure}[htb]
\centering
\includegraphics[width=0.45\textwidth,trim={0 0 0 2cm},clip]{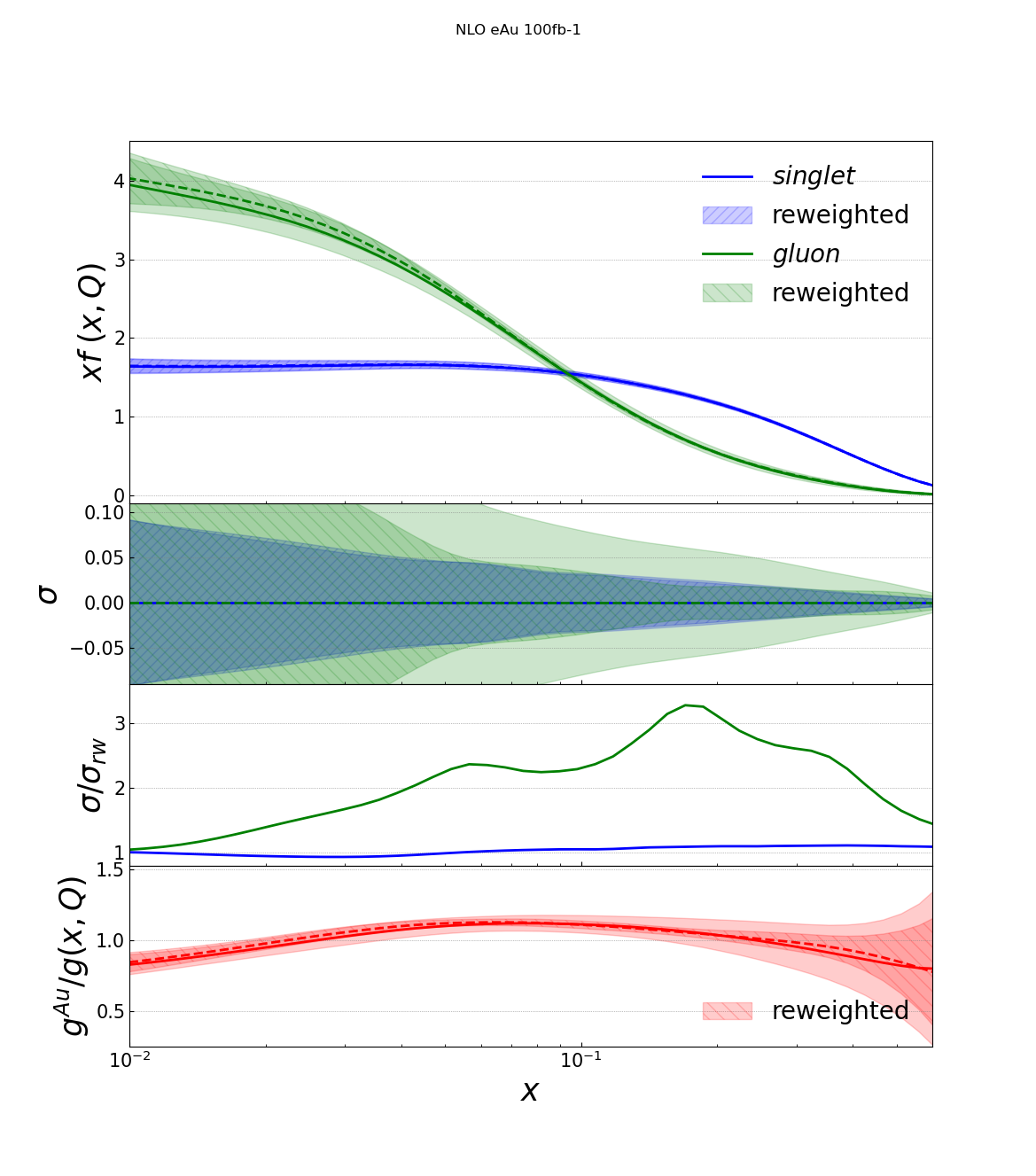}
\caption{Same as Fig.~\ref{fig:CTEQNLO} but for the EPPS~\cite{Eskola:2021nhw} nuclear PDF set with the bottom plot showing the ratio of the nuclear gluon distribution $g^{Au}(x)$ from EPPS~\cite{Eskola:2021nhw} for gold to the respective proton gluon distribution $g(x)$ from CTEQ~\cite{Hou:2019efy}
as a function of the parton momentum $x$. For small $x\sim 10^{-2}$ the shadowing region is visible, while for $x\sim10^{-1}$ the anti-shadowing region is visible.}
\label{fig:EPPS21nlo_CT18Anlo_Au197}
\end{figure}

For our impact study, we focus on the latest PDF set of the EPPS family~\cite{Eskola:2021nhw}. Through the reweighting procedure, the effective set of replicas is reduced from 1000 to approximately 108. Figure~\ref{fig:EPPS21nlo_CT18Anlo_Au197} illustrates the impact on the singlet and gluon PDF for a gold target. Similar to the proton case, we observe a mild impact on the singlet PDF, while the gluon PDF exhibits a significant influence.

In the bottom plot of Fig.~\ref{fig:EPPS21nlo_CT18Anlo_Au197}, we directly compare the impact on the proton and nuclear gluon PDF by plotting the respective ratio. Notably, we observe a substantial reduction in uncertainties in the anti-shadowing region around $x\sim 0.1$. This finding highlights the pronounced effect of the measurements on both the proton and nuclear gluon PDFs.

\subsection{Proton gluon helicity PDF}

With longitudinally polarized electron and proton beams, one can perform double spin asymmetry $A_{\rm LL}$ measurements in 
the $\vec{e}+\vec{p}\to e' + D^0 + X$ process to access the gluon helicity distribution:
\begin{equation}
\begin{aligned}
    A_{LL}^{\vec{e}+\vec{p} \to e' + D^0 + X}
    &= \frac{1}{P_e P_p}\frac{N^{++}-N^{+-}}{N^{++}+N^{+-}} \\
    &\approx\frac{y(2-y)}{y^2+2(1-y)}\frac{g_1^c(x,Q^2)}{F_1^c(x,Q^2)} \\
    &=\frac{y(2-y)}{y^2+2(1-y)}A_1^{c},
\label{eq:A1}
\end{aligned}
\end{equation}
where $N^{++}$ and $N^{+-}$ are the counts normalized by luminosity for different beam helicity states respectively and $P_e$ ($P_p$) is the electron (proton) beam
polarization.
The $A_1^c$ observable provides direct access to the $\Delta g/g$ ratio at leading order. The COMPASS collaboration conducted a pioneering measurement in polarized $\mu$-proton collisions~\cite{Adolph:2012ca}. In the era of the EIC, this measurement can be significantly enhanced. First, the presence of a high-quality vertex detector enables thorough studies of $D^0$ decay topology, leading to improved significance in $D^0$ reconstruction. Second, the combination of high luminosity and large acceptance allows for precise measurements across a wide range of kinematics. This has recently been investigated using an all-silicon tracking system at the EIC~\cite{Anderle:2021hpa}. With a lower center-of-mass energy, the EicC will extend the measurement into the relatively high $x$ region. It is expected that in this region, the ratio of polarized and unpolarized gluons, $\Delta g/g$, will exhibit significantly larger values compared to the small $x$ region \cite{Khan:2022vot,Xu:2022abw}.

\begin{figure}[H]
\centering
\includegraphics[width=0.49\textwidth]{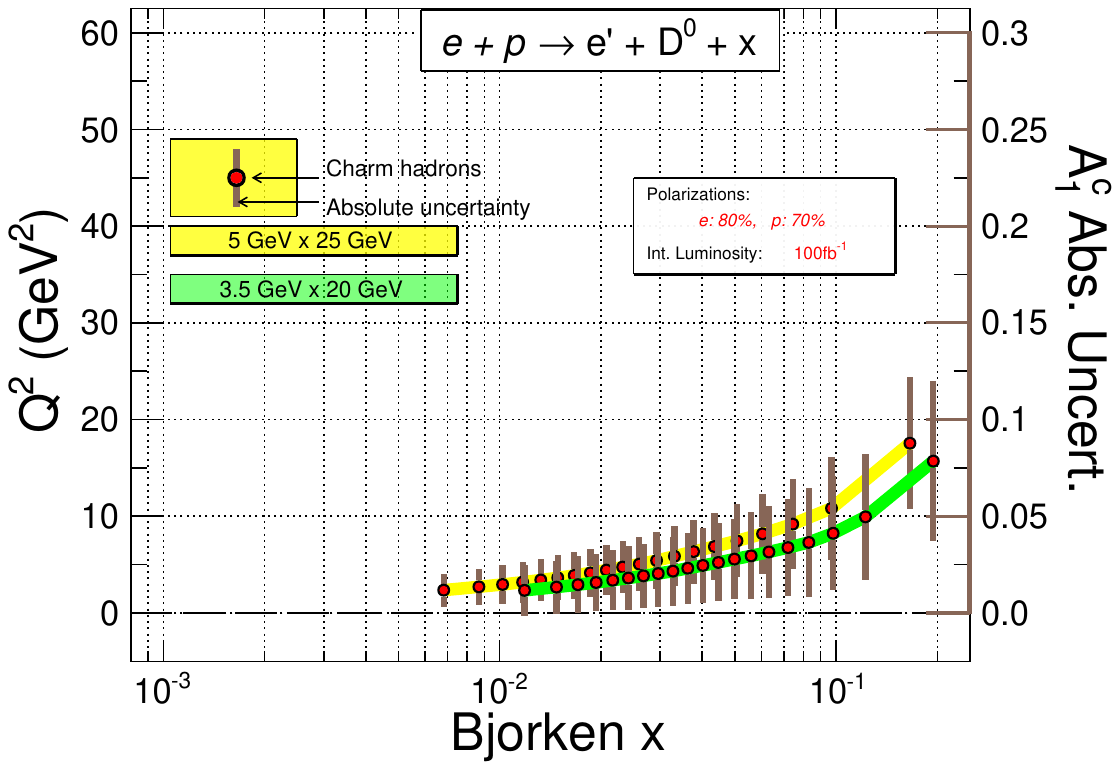}
\caption{Projections of the A$_{1}^{c}$ (formula \ref{eq:A1}) in the
$\vec{e}+\vec{p} \to e + D^0 + X$ process in bins of Bjorken-$x$ for two beam-energy configurations.
The position of each data point in the plot is defined by the
weighted center of Bjorken-$x$ and Q$^2$ for each particular bin. 
The uncertainty indicated for each data point should be interpreted 
using the scale shown on the right-side vertical axis of the plot.
}
\label{fig:A1c}
\end{figure}

In practice, the uncertainty of $A_1^c$ is determined within each Bjorken-$x$ bin based on the number of $D^0$ particles extracted through a fit to the $K\pi$ invariant mass spectrum. The projected uncertainties are displayed in Fig.~\ref{fig:A1c} for two collision energies: 3.5 GeV $\times$ 20 GeV and 5 GeV $\times$ 25 GeV. The mean Bjorken-$x$ and $Q^2$ values for each data point correspond to the positions on the respective axes, while the magnitude of the error is represented by the scale on the right-side vertical axis. The integrated luminosity for each collision energy is set to 100 fb$^{-1}$. The assumed beam polarizations are 80\% for the electron beam and 70\% for the proton beam.

\begin{figure}[htb]
\centering
\includegraphics[width=0.4\textwidth,trim={0.5cm 0 0 2cm},clip]{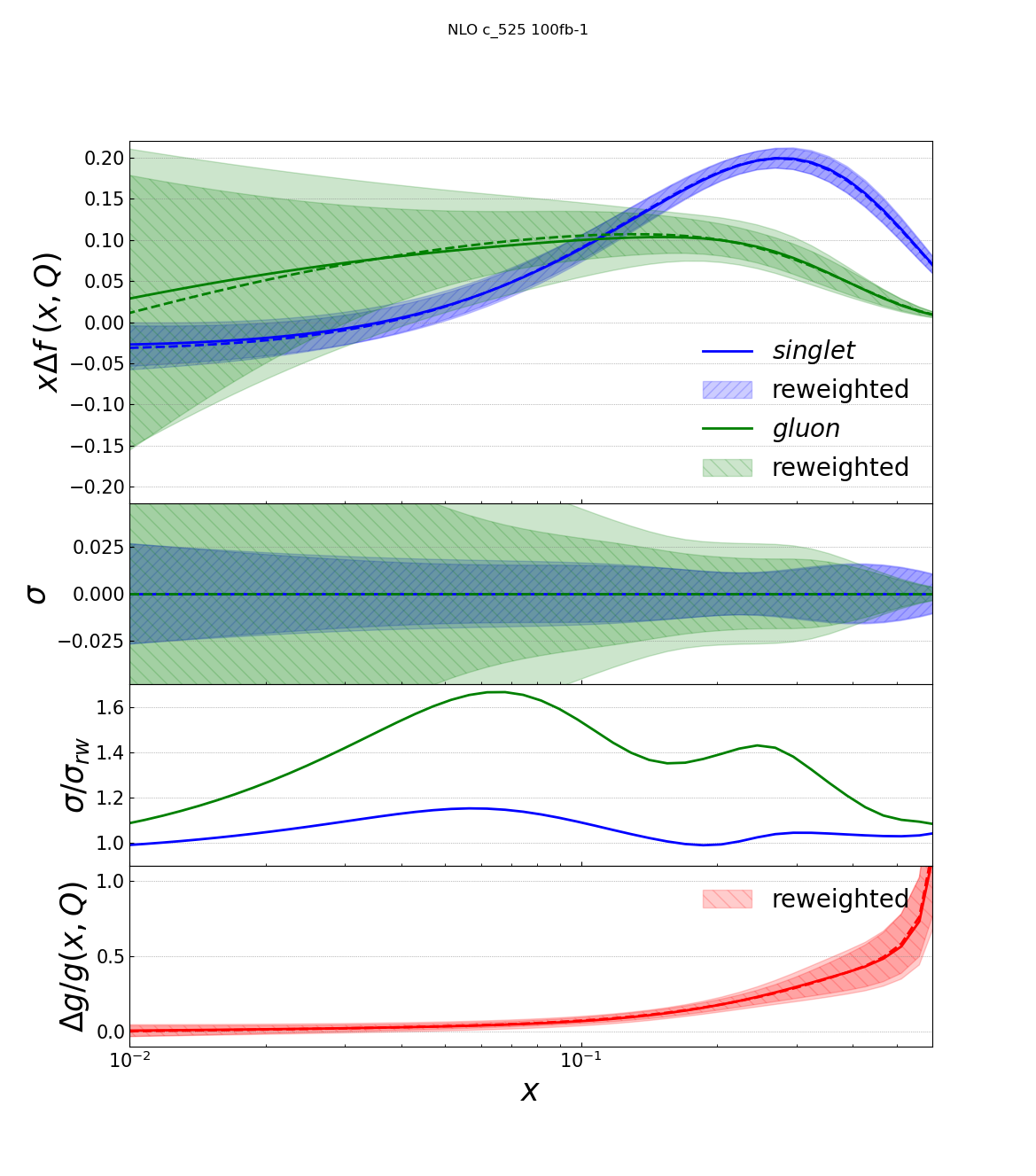} 
\includegraphics[width=0.4\textwidth,trim={0.5cm 0 0 2cm},clip]{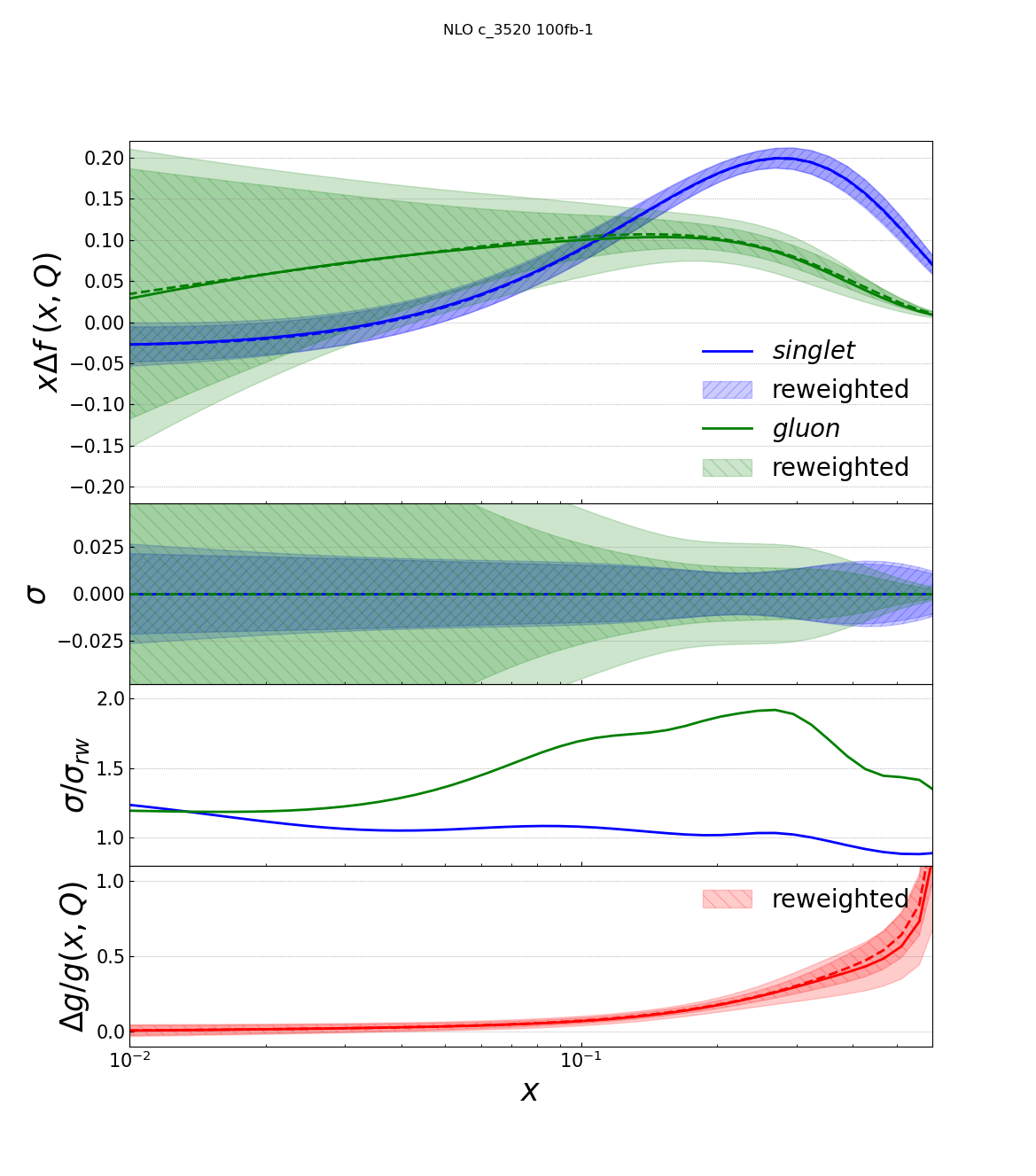}
\caption{Same as Fig.~\ref{fig:CTEQNLO} but for the NNPDFpol1.1 polarized PDF~\cite{Nocera:2014gqa} set for the collision energies 5 GeV $\times$ 25 GeV (top) and 3.5 GeV $\times$ 20 GeV (bottom). The bottom plots show the impact on the ratio of $\Delta g(x)/g(x)$ plotted as a function of the momentum fraction $x$. To produce the ratio we use the central value of NNPDF2.3~\cite{Ball:2012cx} for the unpolarized gluon distributions.}
\label{fig:NNPDFpolNLO}
\end{figure}

The reweighting study is based on the NNPDFpol1.1~\cite{Nocera:2014gqa} helicity distribution set. Fig.~\ref{fig:NNPDFpolNLO} illustrates the impact on the singlet and gluon helicity distributions for the two different energy configurations. The reweighting procedure yields effective replicas of approximately 63 and 49 for the high and low energy configurations, respectively. In both cases, we observe a moderate reduction in the uncertainties of the singlet distributions and a substantial reduction in the uncertainties of the gluon helicity distributions. Notably, the lower energy configuration enhances the precision of the gluon helicity distribution in the relatively high $x$ region, where the signal significance is greater, as depicted in the bottom panel of the plot.

\begin{figure}[htb]
\centering
\includegraphics[width=0.45\textwidth]{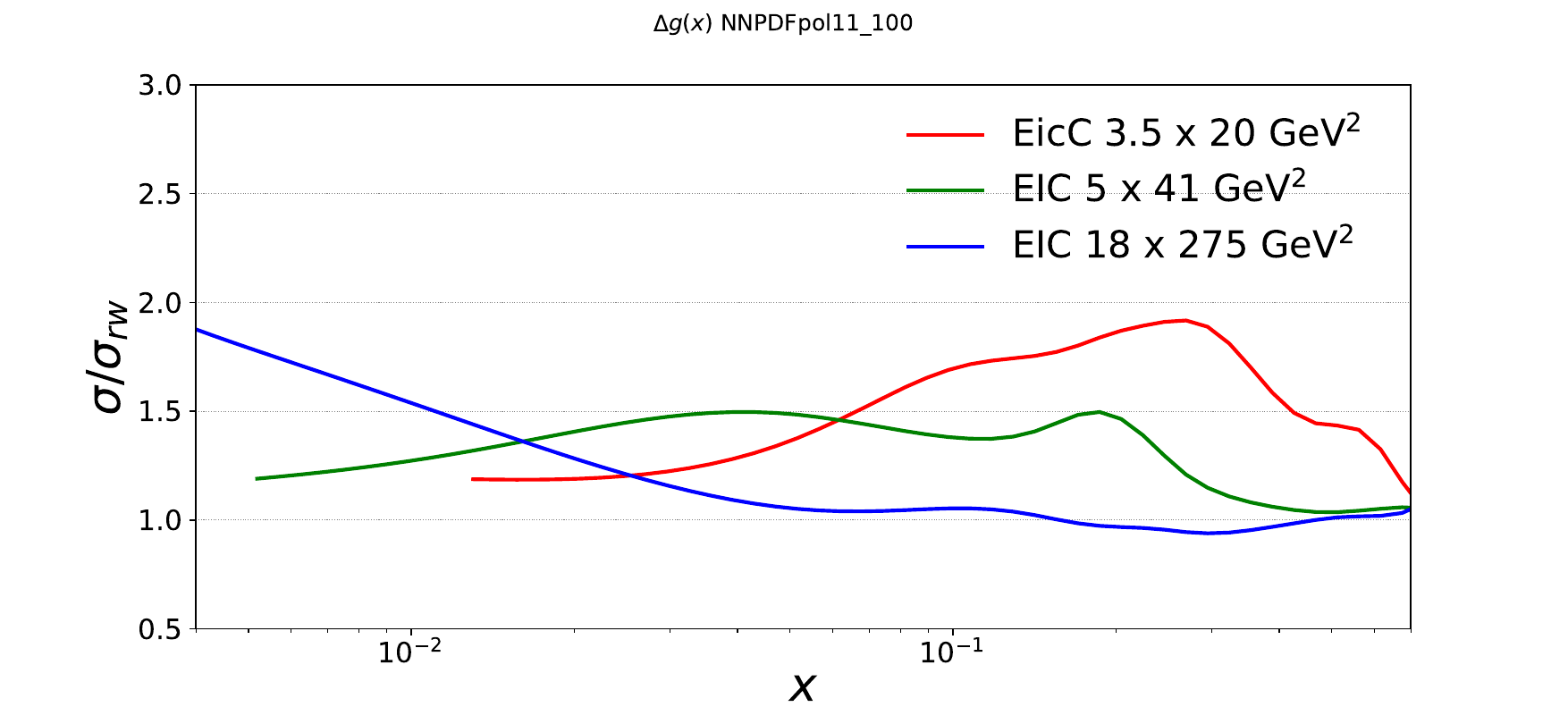}%
\caption{Comparison of ratios between the absolute uncertainties for the gulon helicity distribution before and after reweighting for three configuration energies: two from the EIC as studied in~\cite{Anderle:2021hpa}, namely for the collision energies 5 GeV $\times$ 41 GeV (green) and 18 GeV $\times$ 275 GeV (blue), and one from the EicC energy configuration studied in this paper, 3.5 GeV $\times$ 20 GeV (red). }
\label{fig:NNPDFpolNLO_EIC}
\end{figure}

To demonstrate the dependence of the constraining power on the energy configuration, we compare the impact of the EIC~\cite{Anderle:2021hpa} and the present study in Fig.~\ref{fig:NNPDFpolNLO_EIC}. The plot showcases the potential uncertainty reduction as a function of momentum fraction for the two energy configurations at the EIC and the EicC configuration of a 3.5 GeV $\times$ 20 GeV collision.

\section{Summary}
\label{sec:summary}

In summary, this study presented a conceptual design of a tracking system at the EicC, consisting of silicon and MPGD components. The performance of this system was evaluated through Geant4 simulations. A feasibility study focusing on the reconstruction of $D^0/\bar{D}^0$ mesons was conducted, leveraging the high resolution and excellent vertex reconstruction capabilities of the tracking system. The results indicate that the EicC, as a lower center of mass energy electron-ion collider with high luminosity, offers promising prospects for probing gluon distributions in the relatively high $x$ region by utilizing the abundant $D^0$ production. Specifically, in cases like the determination of $\Delta g/g$, the EicC has an advantage in terms of its extended reach in $x$ where the signal is more prominent. It is important to note that the EicC encompasses a wide range of physics programs, and ongoing efforts involve full simulations and optimizations of the detector system to meet the requirements of various physics channels across diverse kinematic ranges. Although the tracking system design may undergo future revisions, the stringent requirements for heavy flavor measurements, including charm hadron reconstructions, will be upheld at the EicC.

\section*{Acknowledgements}
The authors are grateful for the support from Professor Nu Xu. We also thank Professor Feng Yuan for his helpful suggestions on this work. This work is supported in part by the Strategic Priority Research Program of Chinese Academy of Sciences, under grant number XDB34030301; the National Natural Science Foundation of China (U203220, 12222512) and 100 Talents Program of CAS (E129642Y).
D.~A. and H. X.\ is supported by the Guangdong Major Project of Basic and Applied Basic Research No. 2020B0301030008, No. 2022A1515010683, by the National Natural Science Foundation of China under Grant No. 12022512, No. 12035007.
F.~H.\ is  supported  by  the  European  Research  Council under  the  European  Unions  Horizon  2020  research  and innovation Programme (grant agreement number 740006).

\bibliography{bib.bib}

\begin{thebibliography}{35}%
\makeatletter
\providecommand \@ifxundefined [1]{%
 \@ifx{#1\undefined}
}%
\providecommand \@ifnum [1]{%
 \ifnum #1\expandafter \@firstoftwo
 \else \expandafter \@secondoftwo
 \fi
}%
\providecommand \@ifx [1]{%
 \ifx #1\expandafter \@firstoftwo
 \else \expandafter \@secondoftwo
 \fi
}%
\providecommand \natexlab [1]{#1}%
\providecommand \enquote  [1]{``#1''}%
\providecommand \bibnamefont  [1]{#1}%
\providecommand \bibfnamefont [1]{#1}%
\providecommand \citenamefont [1]{#1}%
\providecommand \href@noop [0]{\@secondoftwo}%
\providecommand \href [0]{\begingroup \@sanitize@url \@href}%
\providecommand \@href[1]{\@@startlink{#1}\@@href}%
\providecommand \@@href[1]{\endgroup#1\@@endlink}%
\providecommand \@sanitize@url [0]{\catcode `\\12\catcode `\$12\catcode
  `\&12\catcode `\#12\catcode `\^12\catcode `\_12\catcode `\%12\relax}%
\providecommand \@@startlink[1]{}%
\providecommand \@@endlink[0]{}%
\providecommand \url  [0]{\begingroup\@sanitize@url \@url }%
\providecommand \@url [1]{\endgroup\@href {#1}{\urlprefix }}%
\providecommand \urlprefix  [0]{URL }%
\providecommand \Eprint [0]{\href }%
\providecommand \doibase [0]{http://dx.doi.org/}%
\providecommand \selectlanguage [0]{\@gobble}%
\providecommand \bibinfo  [0]{\@secondoftwo}%
\providecommand \bibfield  [0]{\@secondoftwo}%
\providecommand \translation [1]{[#1]}%
\providecommand \BibitemOpen [0]{}%
\providecommand \bibitemStop [0]{}%
\providecommand \bibitemNoStop [0]{.\EOS\space}%
\providecommand \EOS [0]{\spacefactor3000\relax}%
\providecommand \BibitemShut  [1]{\csname bibitem#1\endcsname}%
\let\auto@bib@innerbib\@empty
\bibitem [{\citenamefont {Anderle}\ \emph
  {et~al.}(2021{\natexlab{a}})\citenamefont {Anderle} \emph
  {et~al.}}]{Anderle:2021wcy}%
  \BibitemOpen
  \bibfield  {author} {\bibinfo {author} {\bibfnamefont {D.~P.}\ \bibnamefont
  {Anderle}} \emph {et~al.},\ }\href {\doibase 10.1007/s11467-021-1062-0}
  {\bibfield  {journal} {\bibinfo  {journal} {Front. Phys. (Beijing)}\ }\textbf
  {\bibinfo {volume} {16}},\ \bibinfo {pages} {64701} (\bibinfo {year}
  {2021}{\natexlab{a}})},\ \Eprint {http://arxiv.org/abs/2102.09222}
  {arXiv:2102.09222 [nucl-ex]} \BibitemShut {NoStop}%
\bibitem [{\citenamefont {Kelsey}\ \emph {et~al.}(2021)\citenamefont {Kelsey},
  \citenamefont {Cruz-Torres}, \citenamefont {Dong}, \citenamefont {Ji},
  \citenamefont {Radhakrishnan},\ and\ \citenamefont
  {Sichtermann}}]{Kelsey:2021gpk}%
  \BibitemOpen
  \bibfield  {author} {\bibinfo {author} {\bibfnamefont {M.}~\bibnamefont
  {Kelsey}}, \bibinfo {author} {\bibfnamefont {R.}~\bibnamefont {Cruz-Torres}},
  \bibinfo {author} {\bibfnamefont {X.}~\bibnamefont {Dong}}, \bibinfo {author}
  {\bibfnamefont {Y.}~\bibnamefont {Ji}}, \bibinfo {author} {\bibfnamefont
  {S.}~\bibnamefont {Radhakrishnan}}, \ and\ \bibinfo {author} {\bibfnamefont
  {E.}~\bibnamefont {Sichtermann}},\ }\href {\doibase
  10.1103/PhysRevD.104.054002} {\bibfield  {journal} {\bibinfo  {journal}
  {Phys. Rev. D}\ }\textbf {\bibinfo {volume} {104}},\ \bibinfo {pages}
  {054002} (\bibinfo {year} {2021})},\ \Eprint
  {http://arxiv.org/abs/2107.05632} {arXiv:2107.05632 [hep-ph]} \BibitemShut
  {NoStop}%
\bibitem [{\citenamefont {Anderle}\ \emph
  {et~al.}(2021{\natexlab{b}})\citenamefont {Anderle}, \citenamefont {Dong},
  \citenamefont {Hekhorn}, \citenamefont {Kelsey}, \citenamefont
  {Radhakrishnan}, \citenamefont {Sichtermann}, \citenamefont {Xia},
  \citenamefont {Xing}, \citenamefont {Yuan},\ and\ \citenamefont
  {Zhao}}]{Anderle:2021hpa}%
  \BibitemOpen
  \bibfield  {author} {\bibinfo {author} {\bibfnamefont {D.~P.}\ \bibnamefont
  {Anderle}}, \bibinfo {author} {\bibfnamefont {X.}~\bibnamefont {Dong}},
  \bibinfo {author} {\bibfnamefont {F.}~\bibnamefont {Hekhorn}}, \bibinfo
  {author} {\bibfnamefont {M.}~\bibnamefont {Kelsey}}, \bibinfo {author}
  {\bibfnamefont {S.}~\bibnamefont {Radhakrishnan}}, \bibinfo {author}
  {\bibfnamefont {E.}~\bibnamefont {Sichtermann}}, \bibinfo {author}
  {\bibfnamefont {L.}~\bibnamefont {Xia}}, \bibinfo {author} {\bibfnamefont
  {H.}~\bibnamefont {Xing}}, \bibinfo {author} {\bibfnamefont {F.}~\bibnamefont
  {Yuan}}, \ and\ \bibinfo {author} {\bibfnamefont {Y.}~\bibnamefont {Zhao}},\
  }\href {\doibase 10.1103/PhysRevD.104.114039} {\bibfield  {journal} {\bibinfo
   {journal} {Phys. Rev. D}\ }\textbf {\bibinfo {volume} {104}},\ \bibinfo
  {pages} {114039} (\bibinfo {year} {2021}{\natexlab{b}})},\ \Eprint
  {http://arxiv.org/abs/2110.04489} {arXiv:2110.04489 [hep-ex]} \BibitemShut
  {NoStop}%
\bibitem [{\citenamefont {Dong}\ \emph {et~al.}(2022)\citenamefont {Dong},
  \citenamefont {Ji}, \citenamefont {Kelsey}, \citenamefont {Radhakrishnan},
  \citenamefont {Sichtermann},\ and\ \citenamefont {Zhao}}]{Dong:2022xbd}%
  \BibitemOpen
  \bibfield  {author} {\bibinfo {author} {\bibfnamefont {X.}~\bibnamefont
  {Dong}}, \bibinfo {author} {\bibfnamefont {Y.}~\bibnamefont {Ji}}, \bibinfo
  {author} {\bibfnamefont {M.}~\bibnamefont {Kelsey}}, \bibinfo {author}
  {\bibfnamefont {S.}~\bibnamefont {Radhakrishnan}}, \bibinfo {author}
  {\bibfnamefont {E.}~\bibnamefont {Sichtermann}}, \ and\ \bibinfo {author}
  {\bibfnamefont {Y.}~\bibnamefont {Zhao}},\ }\href@noop {} {\  (\bibinfo
  {year} {2022})},\ \Eprint {http://arxiv.org/abs/2210.08609} {arXiv:2210.08609
  [hep-ph]} \BibitemShut {NoStop}%
\bibitem [{\citenamefont {Abdul~Khalek}\ \emph {et~al.}(2022)\citenamefont
  {Abdul~Khalek} \emph {et~al.}}]{AbdulKhalek:2021gbh}%
  \BibitemOpen
  \bibfield  {author} {\bibinfo {author} {\bibfnamefont {R.}~\bibnamefont
  {Abdul~Khalek}} \emph {et~al.},\ }\href {\doibase
  10.1016/j.nuclphysa.2022.122447} {\bibfield  {journal} {\bibinfo  {journal}
  {Nucl. Phys. A}\ }\textbf {\bibinfo {volume} {1026}},\ \bibinfo {pages}
  {122447} (\bibinfo {year} {2022})},\ \Eprint
  {http://arxiv.org/abs/2103.05419} {arXiv:2103.05419 [physics.ins-det]}
  \BibitemShut {NoStop}%
\bibitem [{\citenamefont {Sjostrand}\ \emph {et~al.}(2006)\citenamefont
  {Sjostrand}, \citenamefont {Mrenna},\ and\ \citenamefont
  {Skands}}]{Sjostrand:2006za}%
  \BibitemOpen
  \bibfield  {author} {\bibinfo {author} {\bibfnamefont {T.}~\bibnamefont
  {Sjostrand}}, \bibinfo {author} {\bibfnamefont {S.}~\bibnamefont {Mrenna}}, \
  and\ \bibinfo {author} {\bibfnamefont {P.~Z.}\ \bibnamefont {Skands}},\
  }\href {\doibase 10.1088/1126-6708/2006/05/026} {\bibfield  {journal}
  {\bibinfo  {journal} {JHEP}\ }\textbf {\bibinfo {volume} {05}},\ \bibinfo
  {pages} {026} (\bibinfo {year} {2006})},\ \Eprint
  {http://arxiv.org/abs/hep-ph/0603175} {arXiv:hep-ph/0603175} \BibitemShut
  {NoStop}%
\bibitem [{\citenamefont {Contin}\ \emph {et~al.}(2018)\citenamefont {Contin}
  \emph {et~al.}}]{Contin:2017mck}%
  \BibitemOpen
  \bibfield  {author} {\bibinfo {author} {\bibfnamefont {G.}~\bibnamefont
  {Contin}} \emph {et~al.},\ }\href {\doibase 10.1016/j.nima.2018.03.003}
  {\bibfield  {journal} {\bibinfo  {journal} {Nucl. Instrum. Meth. A}\ }\textbf
  {\bibinfo {volume} {907}},\ \bibinfo {pages} {60} (\bibinfo {year} {2018})},\
  \Eprint {http://arxiv.org/abs/1710.02176} {arXiv:1710.02176
  [physics.ins-det]} \BibitemShut {NoStop}%
\bibitem [{\citenamefont {Abelev}\ \emph {et~al.}(2014)\citenamefont {Abelev}
  \emph {et~al.}}]{ALICE:2013nwm}%
  \BibitemOpen
  \bibfield  {author} {\bibinfo {author} {\bibfnamefont {B.}~\bibnamefont
  {Abelev}} \emph {et~al.} (\bibinfo {collaboration} {ALICE}),\ }\href
  {\doibase 10.1088/0954-3899/41/8/087002} {\bibfield  {journal} {\bibinfo
  {journal} {J. Phys. G}\ }\textbf {\bibinfo {volume} {41}},\ \bibinfo {pages}
  {087002} (\bibinfo {year} {2014})}\BibitemShut {NoStop}%
\bibitem [{\citenamefont {Aglieri~Rinella}(2017)}]{AglieriRinella:2017lym}%
  \BibitemOpen
  \bibfield  {author} {\bibinfo {author} {\bibfnamefont {G.}~\bibnamefont
  {Aglieri~Rinella}} (\bibinfo {collaboration} {ALICE}),\ }\href {\doibase
  10.1016/j.nima.2016.05.016} {\bibfield  {journal} {\bibinfo  {journal} {Nucl.
  Instrum. Meth. A}\ }\textbf {\bibinfo {volume} {845}},\ \bibinfo {pages}
  {583} (\bibinfo {year} {2017})}\BibitemShut {NoStop}%
\bibitem [{\citenamefont {Musa}(2019)}]{Musa:2703140}%
  \BibitemOpen
  \bibfield  {author} {\bibinfo {author} {\bibfnamefont {L.}~\bibnamefont
  {Musa}},\ }\href {https://cds.cern.ch/record/2703140} {\emph {\bibinfo
  {title} {{Letter of Intent for an ALICE ITS Upgrade in LS3}}}},\ \bibinfo
  {type} {Tech. Rep.}\ (\bibinfo  {institution} {CERN},\ \bibinfo {address}
  {Geneva},\ \bibinfo {year} {2019})\BibitemShut {NoStop}%
\bibitem [{\citenamefont {Sauli}(1997)}]{Sauli:1997qp}%
  \BibitemOpen
  \bibfield  {author} {\bibinfo {author} {\bibfnamefont {F.}~\bibnamefont
  {Sauli}},\ }\href {\doibase 10.1016/S0168-9002(96)01172-2} {\bibfield
  {journal} {\bibinfo  {journal} {Nucl. Instrum. Meth. A}\ }\textbf {\bibinfo
  {volume} {386}},\ \bibinfo {pages} {531} (\bibinfo {year}
  {1997})}\BibitemShut {NoStop}%
\bibitem [{\citenamefont {Giomataris}\ \emph {et~al.}(1996)\citenamefont
  {Giomataris}, \citenamefont {Rebourgeard}, \citenamefont {Robert},\ and\
  \citenamefont {Charpak}}]{Giomataris:1995fq}%
  \BibitemOpen
  \bibfield  {author} {\bibinfo {author} {\bibfnamefont {Y.}~\bibnamefont
  {Giomataris}}, \bibinfo {author} {\bibfnamefont {P.}~\bibnamefont
  {Rebourgeard}}, \bibinfo {author} {\bibfnamefont {J.~P.}\ \bibnamefont
  {Robert}}, \ and\ \bibinfo {author} {\bibfnamefont {G.}~\bibnamefont
  {Charpak}},\ }\href {\doibase 10.1016/0168-9002(96)00175-1} {\bibfield
  {journal} {\bibinfo  {journal} {Nucl. Instrum. Meth. A}\ }\textbf {\bibinfo
  {volume} {376}},\ \bibinfo {pages} {29} (\bibinfo {year} {1996})}\BibitemShut
  {NoStop}%
\bibitem [{\citenamefont {Bencivenni}\ \emph {et~al.}(2015)\citenamefont
  {Bencivenni}, \citenamefont {De~Oliveira}, \citenamefont {Morello},\ and\
  \citenamefont {Lener}}]{Bencivenni:2014exa}%
  \BibitemOpen
  \bibfield  {author} {\bibinfo {author} {\bibfnamefont {G.}~\bibnamefont
  {Bencivenni}}, \bibinfo {author} {\bibfnamefont {R.}~\bibnamefont
  {De~Oliveira}}, \bibinfo {author} {\bibfnamefont {G.}~\bibnamefont
  {Morello}}, \ and\ \bibinfo {author} {\bibfnamefont {M.~P.}\ \bibnamefont
  {Lener}},\ }\href {\doibase 10.1088/1748-0221/10/02/P02008} {\bibfield
  {journal} {\bibinfo  {journal} {JINST}\ }\textbf {\bibinfo {volume} {10}},\
  \bibinfo {pages} {P02008} (\bibinfo {year} {2015})},\ \Eprint
  {http://arxiv.org/abs/1411.2466} {arXiv:1411.2466 [physics.ins-det]}
  \BibitemShut {NoStop}%
\bibitem [{\citenamefont {Al-Turany}\ \emph {et~al.}(2012)\citenamefont
  {Al-Turany}, \citenamefont {Bertini}, \citenamefont {Karabowicz},
  \citenamefont {Kresan}, \citenamefont {Malzacher}, \citenamefont
  {Stockmanns},\ and\ \citenamefont {Uhlig}}]{Al-Turany:2012zfk}%
  \BibitemOpen
  \bibfield  {author} {\bibinfo {author} {\bibfnamefont {M.}~\bibnamefont
  {Al-Turany}}, \bibinfo {author} {\bibfnamefont {D.}~\bibnamefont {Bertini}},
  \bibinfo {author} {\bibfnamefont {R.}~\bibnamefont {Karabowicz}}, \bibinfo
  {author} {\bibfnamefont {D.}~\bibnamefont {Kresan}}, \bibinfo {author}
  {\bibfnamefont {P.}~\bibnamefont {Malzacher}}, \bibinfo {author}
  {\bibfnamefont {T.}~\bibnamefont {Stockmanns}}, \ and\ \bibinfo {author}
  {\bibfnamefont {F.}~\bibnamefont {Uhlig}},\ }\href {\doibase
  10.1088/1742-6596/396/2/022001} {\bibfield  {journal} {\bibinfo  {journal}
  {J. Phys. Conf. Ser.}\ }\textbf {\bibinfo {volume} {396}},\ \bibinfo {pages}
  {022001} (\bibinfo {year} {2012})}\BibitemShut {NoStop}%
\bibitem [{\citenamefont {Agostinelli}\ \emph {et~al.}(2003)\citenamefont
  {Agostinelli} \emph {et~al.}}]{GEANT4:2002zbu}%
  \BibitemOpen
  \bibfield  {author} {\bibinfo {author} {\bibfnamefont {S.}~\bibnamefont
  {Agostinelli}} \emph {et~al.} (\bibinfo {collaboration} {GEANT4}),\ }\href
  {\doibase 10.1016/S0168-9002(03)01368-8} {\bibfield  {journal} {\bibinfo
  {journal} {Nucl. Instrum. Meth. A}\ }\textbf {\bibinfo {volume} {506}},\
  \bibinfo {pages} {250} (\bibinfo {year} {2003})}\BibitemShut {NoStop}%
\bibitem [{\citenamefont {Rauch}\ and\ \citenamefont
  {Schl\"uter}(2015)}]{Rauch:2014wta}%
  \BibitemOpen
  \bibfield  {author} {\bibinfo {author} {\bibfnamefont {J.}~\bibnamefont
  {Rauch}}\ and\ \bibinfo {author} {\bibfnamefont {T.}~\bibnamefont
  {Schl\"uter}},\ }\href {\doibase 10.1088/1742-6596/608/1/012042} {\bibfield
  {journal} {\bibinfo  {journal} {J. Phys. Conf. Ser.}\ }\textbf {\bibinfo
  {volume} {608}},\ \bibinfo {pages} {012042} (\bibinfo {year} {2015})},\
  \Eprint {http://arxiv.org/abs/1410.3698} {arXiv:1410.3698 [physics.ins-det]}
  \BibitemShut {NoStop}%
\bibitem [{\citenamefont {Waltenberger}(2011)}]{Waltenberger:2011zz}%
  \BibitemOpen
  \bibfield  {author} {\bibinfo {author} {\bibfnamefont {W.}~\bibnamefont
  {Waltenberger}},\ }\href {\doibase 10.1109/TNS.2011.2119492} {\bibfield
  {journal} {\bibinfo  {journal} {IEEE Trans. Nucl. Sci.}\ }\textbf {\bibinfo
  {volume} {58}},\ \bibinfo {pages} {434} (\bibinfo {year} {2011})}\BibitemShut
  {NoStop}%
\bibitem [{\citenamefont {Giele}\ and\ \citenamefont
  {Keller}(1998)}]{Giele:1998gw}%
  \BibitemOpen
  \bibfield  {author} {\bibinfo {author} {\bibfnamefont {W.~T.}\ \bibnamefont
  {Giele}}\ and\ \bibinfo {author} {\bibfnamefont {S.}~\bibnamefont {Keller}},\
  }\href {\doibase 10.1103/PhysRevD.58.094023} {\bibfield  {journal} {\bibinfo
  {journal} {Phys. Rev. D}\ }\textbf {\bibinfo {volume} {58}},\ \bibinfo
  {pages} {094023} (\bibinfo {year} {1998})},\ \Eprint
  {http://arxiv.org/abs/hep-ph/9803393} {arXiv:hep-ph/9803393} \BibitemShut
  {NoStop}%
\bibitem [{\citenamefont {Ball}\ \emph {et~al.}(2011)\citenamefont {Ball},
  \citenamefont {Bertone}, \citenamefont {Cerutti}, \citenamefont {Del~Debbio},
  \citenamefont {Forte}, \citenamefont {Guffanti}, \citenamefont {Latorre},
  \citenamefont {Rojo},\ and\ \citenamefont {Ubiali}}]{Ball:2010gb}%
  \BibitemOpen
  \bibfield  {author} {\bibinfo {author} {\bibfnamefont {R.~D.}\ \bibnamefont
  {Ball}}, \bibinfo {author} {\bibfnamefont {V.}~\bibnamefont {Bertone}},
  \bibinfo {author} {\bibfnamefont {F.}~\bibnamefont {Cerutti}}, \bibinfo
  {author} {\bibfnamefont {L.}~\bibnamefont {Del~Debbio}}, \bibinfo {author}
  {\bibfnamefont {S.}~\bibnamefont {Forte}}, \bibinfo {author} {\bibfnamefont
  {A.}~\bibnamefont {Guffanti}}, \bibinfo {author} {\bibfnamefont {J.~I.}\
  \bibnamefont {Latorre}}, \bibinfo {author} {\bibfnamefont {J.}~\bibnamefont
  {Rojo}}, \ and\ \bibinfo {author} {\bibfnamefont {M.}~\bibnamefont {Ubiali}}
  (\bibinfo {collaboration} {NNPDF}),\ }\href {\doibase
  10.1016/j.nuclphysb.2011.03.017} {\bibfield  {journal} {\bibinfo  {journal}
  {Nucl. Phys. B}\ }\textbf {\bibinfo {volume} {849}},\ \bibinfo {pages} {112}
  (\bibinfo {year} {2011})},\ \bibinfo {note} {[Erratum: Nucl.Phys.B 854,
  926--927 (2012), Erratum: Nucl.Phys.B 855, 927--928 (2012)]},\ \Eprint
  {http://arxiv.org/abs/1012.0836} {arXiv:1012.0836 [hep-ph]} \BibitemShut
  {NoStop}%
\bibitem [{\citenamefont {Ball}\ \emph {et~al.}(2012)\citenamefont {Ball},
  \citenamefont {Bertone}, \citenamefont {Cerutti}, \citenamefont {Del~Debbio},
  \citenamefont {Forte}, \citenamefont {Guffanti}, \citenamefont {Hartland},
  \citenamefont {Latorre}, \citenamefont {Rojo},\ and\ \citenamefont
  {Ubiali}}]{Ball:2011gg}%
  \BibitemOpen
  \bibfield  {author} {\bibinfo {author} {\bibfnamefont {R.~D.}\ \bibnamefont
  {Ball}}, \bibinfo {author} {\bibfnamefont {V.}~\bibnamefont {Bertone}},
  \bibinfo {author} {\bibfnamefont {F.}~\bibnamefont {Cerutti}}, \bibinfo
  {author} {\bibfnamefont {L.}~\bibnamefont {Del~Debbio}}, \bibinfo {author}
  {\bibfnamefont {S.}~\bibnamefont {Forte}}, \bibinfo {author} {\bibfnamefont
  {A.}~\bibnamefont {Guffanti}}, \bibinfo {author} {\bibfnamefont {N.~P.}\
  \bibnamefont {Hartland}}, \bibinfo {author} {\bibfnamefont {J.~I.}\
  \bibnamefont {Latorre}}, \bibinfo {author} {\bibfnamefont {J.}~\bibnamefont
  {Rojo}}, \ and\ \bibinfo {author} {\bibfnamefont {M.}~\bibnamefont
  {Ubiali}},\ }\href {\doibase 10.1016/j.nuclphysb.2011.10.018} {\bibfield
  {journal} {\bibinfo  {journal} {Nucl. Phys. B}\ }\textbf {\bibinfo {volume}
  {855}},\ \bibinfo {pages} {608} (\bibinfo {year} {2012})},\ \Eprint
  {http://arxiv.org/abs/1108.1758} {arXiv:1108.1758 [hep-ph]} \BibitemShut
  {NoStop}%
\bibitem [{pyt()}]{pythiaerhic}%
  \BibitemOpen
  \href@noop {} {\enquote {\bibinfo {title} {{PYTHIA6 with Radiative
  Corrections}},}\ }\bibinfo {howpublished}
  {\url{https://eic.github.io/software/pythia6.html}}\BibitemShut {NoStop}%
\bibitem [{\citenamefont {Rosenbluth}(1950)}]{PhysRev.79.615}%
  \BibitemOpen
  \bibfield  {author} {\bibinfo {author} {\bibfnamefont {M.~N.}\ \bibnamefont
  {Rosenbluth}},\ }\href {\doibase 10.1103/PhysRev.79.615} {\bibfield
  {journal} {\bibinfo  {journal} {Phys. Rev.}\ }\textbf {\bibinfo {volume}
  {79}},\ \bibinfo {pages} {615} (\bibinfo {year} {1950})}\BibitemShut
  {NoStop}%
\bibitem [{\citenamefont {Group}\ \emph {et~al.}(2022)\citenamefont {Group},
  \citenamefont {Workman}, \citenamefont {Burkert} \emph
  {et~al.}}]{10.1093/ptep/ptac097}%
  \BibitemOpen
  \bibfield  {author} {\bibinfo {author} {\bibfnamefont {P.~D.}\ \bibnamefont
  {Group}}, \bibinfo {author} {\bibfnamefont {R.~L.}\ \bibnamefont {Workman}},
  \bibinfo {author} {\bibfnamefont {V.~D.}\ \bibnamefont {Burkert}},  \emph
  {et~al.},\ }\href {\doibase 10.1093/ptep/ptac097} {\bibfield  {journal}
  {\bibinfo  {journal} {Progress of Theoretical and Experimental Physics}\
  }\textbf {\bibinfo {volume} {2022}} (\bibinfo {year} {2022}),\
  10.1093/ptep/ptac097},\ \bibinfo {note} {083C01},\ \Eprint
  {http://arxiv.org/abs/https://academic.oup.com/ptep/article-pdf/2022/8/083C01/45434166/ptac097.pdf}
  {https://academic.oup.com/ptep/article-pdf/2022/8/083C01/45434166/ptac097.pdf}
  \BibitemShut {NoStop}%
\bibitem [{\citenamefont {Sjöstrand}\ \emph {et~al.}(2006)\citenamefont
  {Sjöstrand}, \citenamefont {Mrenna},\ and\ \citenamefont
  {Skands}}]{Sj_strand_2006}%
  \BibitemOpen
  \bibfield  {author} {\bibinfo {author} {\bibfnamefont {T.}~\bibnamefont
  {Sjöstrand}}, \bibinfo {author} {\bibfnamefont {S.}~\bibnamefont {Mrenna}},
  \ and\ \bibinfo {author} {\bibfnamefont {P.}~\bibnamefont {Skands}},\ }\href
  {\doibase 10.1088/1126-6708/2006/05/026} {\bibfield  {journal} {\bibinfo
  {journal} {Journal of High Energy Physics}\ }\textbf {\bibinfo {volume}
  {2006}},\ \bibinfo {pages} {026} (\bibinfo {year} {2006})}\BibitemShut
  {NoStop}%
\bibitem [{\citenamefont {Hou}\ \emph {et~al.}(2021)\citenamefont {Hou} \emph
  {et~al.}}]{Hou:2019efy}%
  \BibitemOpen
  \bibfield  {author} {\bibinfo {author} {\bibfnamefont {T.-J.}\ \bibnamefont
  {Hou}} \emph {et~al.},\ }\href {\doibase 10.1103/PhysRevD.103.014013}
  {\bibfield  {journal} {\bibinfo  {journal} {Phys. Rev. D}\ }\textbf {\bibinfo
  {volume} {103}},\ \bibinfo {pages} {014013} (\bibinfo {year} {2021})},\
  \Eprint {http://arxiv.org/abs/1912.10053} {arXiv:1912.10053 [hep-ph]}
  \BibitemShut {NoStop}%
\bibitem [{\citenamefont {Gao}\ and\ \citenamefont
  {Nadolsky}(2014)}]{Gao:2013bia}%
  \BibitemOpen
  \bibfield  {author} {\bibinfo {author} {\bibfnamefont {J.}~\bibnamefont
  {Gao}}\ and\ \bibinfo {author} {\bibfnamefont {P.}~\bibnamefont {Nadolsky}},\
  }\href {\doibase 10.1007/JHEP07(2014)035} {\bibfield  {journal} {\bibinfo
  {journal} {JHEP}\ }\textbf {\bibinfo {volume} {07}},\ \bibinfo {pages} {035}
  (\bibinfo {year} {2014})},\ \Eprint {http://arxiv.org/abs/1401.0013}
  {arXiv:1401.0013 [hep-ph]} \BibitemShut {NoStop}%
\bibitem [{\citenamefont {Hou}\ \emph {et~al.}(2017)\citenamefont {Hou} \emph
  {et~al.}}]{Hou:2016sho}%
  \BibitemOpen
  \bibfield  {author} {\bibinfo {author} {\bibfnamefont {T.-J.}\ \bibnamefont
  {Hou}} \emph {et~al.},\ }\href {\doibase 10.1007/JHEP03(2017)099} {\bibfield
  {journal} {\bibinfo  {journal} {JHEP}\ }\textbf {\bibinfo {volume} {03}},\
  \bibinfo {pages} {099} (\bibinfo {year} {2017})},\ \Eprint
  {http://arxiv.org/abs/1607.06066} {arXiv:1607.06066 [hep-ph]} \BibitemShut
  {NoStop}%
\bibitem [{\citenamefont {Candido}\ \emph {et~al.}(2022)\citenamefont
  {Candido}, \citenamefont {Hekhorn},\ and\ \citenamefont
  {Magni}}]{candido_alessandro_2022_6285149}%
  \BibitemOpen
  \bibfield  {author} {\bibinfo {author} {\bibfnamefont {A.}~\bibnamefont
  {Candido}}, \bibinfo {author} {\bibfnamefont {F.}~\bibnamefont {Hekhorn}}, \
  and\ \bibinfo {author} {\bibfnamefont {G.}~\bibnamefont {Magni}},\ }\href
  {\doibase 10.5281/zenodo.6285149} {\enquote {\bibinfo {title} {N3pdf/yadism:
  Fonll-b},}\ } (\bibinfo {year} {2022})\BibitemShut {NoStop}%
\bibitem [{\citenamefont {Forte}\ \emph {et~al.}(2010)\citenamefont {Forte},
  \citenamefont {Laenen}, \citenamefont {Nason},\ and\ \citenamefont
  {Rojo}}]{Forte:2010ta}%
  \BibitemOpen
  \bibfield  {author} {\bibinfo {author} {\bibfnamefont {S.}~\bibnamefont
  {Forte}}, \bibinfo {author} {\bibfnamefont {E.}~\bibnamefont {Laenen}},
  \bibinfo {author} {\bibfnamefont {P.}~\bibnamefont {Nason}}, \ and\ \bibinfo
  {author} {\bibfnamefont {J.}~\bibnamefont {Rojo}},\ }\href {\doibase
  10.1016/j.nuclphysb.2010.03.014} {\bibfield  {journal} {\bibinfo  {journal}
  {Nucl. Phys. B}\ }\textbf {\bibinfo {volume} {834}},\ \bibinfo {pages} {116}
  (\bibinfo {year} {2010})},\ \Eprint {http://arxiv.org/abs/1001.2312}
  {arXiv:1001.2312 [hep-ph]} \BibitemShut {NoStop}%
\bibitem [{\citenamefont {Eskola}\ \emph {et~al.}(2022)\citenamefont {Eskola},
  \citenamefont {Paakkinen}, \citenamefont {Paukkunen},\ and\ \citenamefont
  {Salgado}}]{Eskola:2021nhw}%
  \BibitemOpen
  \bibfield  {author} {\bibinfo {author} {\bibfnamefont {K.~J.}\ \bibnamefont
  {Eskola}}, \bibinfo {author} {\bibfnamefont {P.}~\bibnamefont {Paakkinen}},
  \bibinfo {author} {\bibfnamefont {H.}~\bibnamefont {Paukkunen}}, \ and\
  \bibinfo {author} {\bibfnamefont {C.~A.}\ \bibnamefont {Salgado}},\ }\href
  {\doibase 10.1140/epjc/s10052-022-10359-0} {\bibfield  {journal} {\bibinfo
  {journal} {Eur. Phys. J. C}\ }\textbf {\bibinfo {volume} {82}},\ \bibinfo
  {pages} {413} (\bibinfo {year} {2022})},\ \Eprint
  {http://arxiv.org/abs/2112.12462} {arXiv:2112.12462 [hep-ph]} \BibitemShut
  {NoStop}%
\bibitem [{\citenamefont {Adolph}\ \emph {et~al.}(2013)\citenamefont {Adolph}
  \emph {et~al.}}]{Adolph:2012ca}%
  \BibitemOpen
  \bibfield  {author} {\bibinfo {author} {\bibfnamefont {C.}~\bibnamefont
  {Adolph}} \emph {et~al.} (\bibinfo {collaboration} {COMPASS}),\ }\href
  {\doibase 10.1103/PhysRevD.87.052018} {\bibfield  {journal} {\bibinfo
  {journal} {Phys. Rev. D}\ }\textbf {\bibinfo {volume} {87}},\ \bibinfo
  {pages} {052018} (\bibinfo {year} {2013})},\ \Eprint
  {http://arxiv.org/abs/1211.6849} {arXiv:1211.6849 [hep-ex]} \BibitemShut
  {NoStop}%
\bibitem [{\citenamefont {Khan}\ \emph {et~al.}(2022)\citenamefont {Khan},
  \citenamefont {Liu},\ and\ \citenamefont {Sufian}}]{Khan:2022vot}%
  \BibitemOpen
  \bibfield  {author} {\bibinfo {author} {\bibfnamefont {T.}~\bibnamefont
  {Khan}}, \bibinfo {author} {\bibfnamefont {T.}~\bibnamefont {Liu}}, \ and\
  \bibinfo {author} {\bibfnamefont {R.~S.}\ \bibnamefont {Sufian}},\
  }\href@noop {} {\  (\bibinfo {year} {2022})},\ \Eprint
  {http://arxiv.org/abs/2211.15587} {arXiv:2211.15587 [hep-lat]} \BibitemShut
  {NoStop}%
\bibitem [{\citenamefont {Xu}\ \emph {et~al.}(2022)\citenamefont {Xu},
  \citenamefont {Mondal}, \citenamefont {Zhao}, \citenamefont {Li},\ and\
  \citenamefont {Vary}}]{Xu:2022abw}%
  \BibitemOpen
  \bibfield  {author} {\bibinfo {author} {\bibfnamefont {S.}~\bibnamefont
  {Xu}}, \bibinfo {author} {\bibfnamefont {C.}~\bibnamefont {Mondal}}, \bibinfo
  {author} {\bibfnamefont {X.}~\bibnamefont {Zhao}}, \bibinfo {author}
  {\bibfnamefont {Y.}~\bibnamefont {Li}}, \ and\ \bibinfo {author}
  {\bibfnamefont {J.~P.}\ \bibnamefont {Vary}},\ }\href@noop {} {\  (\bibinfo
  {year} {2022})},\ \Eprint {http://arxiv.org/abs/2209.08584} {arXiv:2209.08584
  [hep-ph]} \BibitemShut {NoStop}%
\bibitem [{\citenamefont {Nocera}\ \emph {et~al.}(2014)\citenamefont {Nocera},
  \citenamefont {Ball}, \citenamefont {Forte}, \citenamefont {Ridolfi},\ and\
  \citenamefont {Rojo}}]{Nocera:2014gqa}%
  \BibitemOpen
  \bibfield  {author} {\bibinfo {author} {\bibfnamefont {E.~R.}\ \bibnamefont
  {Nocera}}, \bibinfo {author} {\bibfnamefont {R.~D.}\ \bibnamefont {Ball}},
  \bibinfo {author} {\bibfnamefont {S.}~\bibnamefont {Forte}}, \bibinfo
  {author} {\bibfnamefont {G.}~\bibnamefont {Ridolfi}}, \ and\ \bibinfo
  {author} {\bibfnamefont {J.}~\bibnamefont {Rojo}} (\bibinfo {collaboration}
  {NNPDF}),\ }\href {\doibase 10.1016/j.nuclphysb.2014.08.008} {\bibfield
  {journal} {\bibinfo  {journal} {Nucl. Phys. B}\ }\textbf {\bibinfo {volume}
  {887}},\ \bibinfo {pages} {276} (\bibinfo {year} {2014})},\ \Eprint
  {http://arxiv.org/abs/1406.5539} {arXiv:1406.5539 [hep-ph]} \BibitemShut
  {NoStop}%
\bibitem [{\citenamefont {Ball}\ \emph {et~al.}(2013)\citenamefont {Ball} \emph
  {et~al.}}]{Ball:2012cx}%
  \BibitemOpen
  \bibfield  {author} {\bibinfo {author} {\bibfnamefont {R.~D.}\ \bibnamefont
  {Ball}} \emph {et~al.},\ }\href {\doibase 10.1016/j.nuclphysb.2012.10.003}
  {\bibfield  {journal} {\bibinfo  {journal} {Nucl. Phys. B}\ }\textbf
  {\bibinfo {volume} {867}},\ \bibinfo {pages} {244} (\bibinfo {year}
  {2013})},\ \Eprint {http://arxiv.org/abs/1207.1303} {arXiv:1207.1303
  [hep-ph]} \BibitemShut {NoStop}%
\end{thebibliography}%
\clearpage
\appendix

\end{document}